\documentclass[10pt]{amsart}

\usepackage{amsfonts}

\usepackage{graphicx}
\usepackage{mathrsfs}
\usepackage{amsmath}
\usepackage{setspace}
\newcommand{\halmos}{{\mbox{\, \vspace{3mm}}} \hfill
\mbox{$\Box$}}

\newtheorem{theorem}{Theorem}
\newtheorem{assumption}{Assumption}

\newtheorem{condition}[theorem]{Condition}
\newtheorem{conjecture}[theorem]{Conjecture}
\newtheorem{corollary}[theorem]{Corollary}

\newtheorem{definition}[theorem]{Definition}
\newtheorem{example}[theorem]{Example}

\newtheorem{lemma}[theorem]{Lemma}

\newtheorem{proposition}[theorem]{Proposition}
\newtheorem{remark}[theorem]{Remark}

\usepackage[english]{babel}
\usepackage[left=3cm,right=3cm,top=2.5cm,bottom=3cm]{geometry}

\usepackage{amsmath,amssymb,amsthm,bbm,color,graphics,version}
\usepackage{mathrsfs}

\newcommand{\bearno}{\begin{eqnarray*}}
\newcommand{\enarno}{\end{eqnarray*}}

\newcommand{\vb}{\vspace{3mm}}

\title{Optimal portfolio selection in an It\^{o}-Markov additive market }

\author{Zbigniew Palmowski}
\address{Department of Applied Mathematics, Faculty of Pure and Applied Mathematics, Wroc\l aw University of Science and Technology, Poland}
\email{zbigniew.palmowski@gmail.com}

\author{{\L}ukasz Stettner}
\address{Institute of Mathematics, Polish Academy of Sciences, Poland }
\email{stettner@impan.pl}

\author{Anna Sulima}
\address{Faculty of Management, Computer Science and Finance, Wroc\l aw University of Economics, Poland  }
\email{anna.sulima@wp.pl}

\date{}
\thanks{ Anna Sulima would like to thank prof. Monique Pontier (Universit\'e Paul Sabatier, Toulouse) for her a lot of plenty of helpful suggestions improving this paper.
This paper is supported by the National Science Centre under the grant 2016/23/B/HS4/00566 (2017-2020).}

\begin{document}

\begin{abstract}
We study a portfolio selection problem in a continuous-time It\^{o}-Markov additive market with prices of financial assets described by Markov additive processes which combine L\'{e}vy processes and regime switching models.
Thus the model takes into account two sources of risk: the jump diffusion risk and the regime switching risk. For this reason the market is incomplete. We complete the market by enlarging it with the use of a set of Markovian jump securities, Markovian power-jump securities and impulse regime switching securities. Moreover, we give conditions under which the market is asymptotic-arbitrage-free. We solve the portfolio selection problem in the It\^{o}-Markov additive market for the power utility and the logarithmic utility.

\vspace{3mm}


\end{abstract}

\maketitle

\pagestyle{myheadings} \markboth{\sc Z.\ Palmowski --- {\L}.\ Stettner --- A.\ Sulima} {\sc Optimal portfolio selection for It\^{o}-Markov additive market }

\vspace{1.8cm}

\tableofcontents

\newpage

\section{Introduction}
\text { } \text { } The portfolio selection problem is an important issue in financial mathematics. The problem is to invest an initial wealth in financial assets so as to maximize the expected utility of the terminal wealth.
Markowitz \cite{Ma} pioneered the use of quantitative methods for the optimal portfolio selection problem and developed the mean-variance approach for portfolio optimization. Explicit solutions for the portfolio selection problem in continuous time were first given by Merton \cite{Me1, Me2}.\\
\text { } \text { } Although Merton's approach produces significant theoretical results, it has some shortcomings from the practical perspective. The first is related to the assumption that the dynamics of a risky asset follows a geometric Brownian motion. Many investigations (e.g. Black et al. \cite{BJS}., Merton \cite{Me3}) have suggested that this market model cannot explain some empirical behaviors of financial time series, such as the asymmetry and heavy-tailedness of the distribution of returns of a time-varying conditional volatility. To model this, the stock price driven by a L\'{e}vy process is more suitable. The portfolio selection problem on a L\'{e}vy market was considered by Niu \cite{Niu} and Corcuera et al. \cite{CGNS}.\\
\text { } \text { } The second important assumption in the original Black-Scholes-Merton model is that the coefficients are fixed. However, this seems to be far from reality, especially if one wishes to consider investment problems
under economic uncertainties over a long time period, where structural changes in
macroeconomic conditions may occur several times and cause fundamental changes in investment opportunity sets. Good candidates for modeling such behaviors seem to be Markov modulated models (otherwise called regime switching). In such models, one set of model parameters is in force at a particular time according to the state of the economy at that time. The set of parameters will change to another set when there is a transition in the state of the economy, which is usually described by a Markov chain. Hence, regime switching models can describe structural changes in macroeconomic conditions or different stages of business cycles (see Zhang \cite{Zha}). Hamilton \cite{Ham} pioneered econometric applications of regime switching models. These models have diverse applications in finance (see Buffington and Elliott \cite{BEl}, Di Masi et al. \cite{DMKRu}, Elliott et al. \cite{ECSi, EHJa, EMTs, ER}, Goldfeld and Quandt \cite{GQu}, Guo \cite{Guo}, Naik \cite{Naik} and Tong \cite{T1, T2}).\\
\text { } \text { } In this paper we consider a market with the prices of financial assets described by It\^{o}-Markov additive processes, which combine L\'{e}vy processes and regime switching models. Such a process evolves as an It\^{o}-L\'evy process
between changes of states of a Markov chain, that is, its parameters depend on the current state of the Markov chain.
In addition, a transition of the Markov chain from state $i$ to state $j$ triggers an additional jump. The use of It\^{o}-Markov additive processes is widespread, making them a classical model in applied probability with a variety of application areas, such as queues, insurance risk, inventories, data communication, finance, environmental problems and others (see Asmussen \cite{A2}, Asmussen and  Albrecher \cite{A1}, Asmussen et al. \cite{asm_avram_pist}, Asmussen and Kella \cite{AK}, Cinlar \cite{Ca, Cb}, Ezhov and Skorokhod \cite{ES}, Ivanovs and  Palmowski \cite{meiv}, Pacheco and Prabhu \cite{PP}, Pacheco et al. \cite{PTP}, Palmowski and Rolski \cite{PR}, Prabhu \cite[Ch. 7]{prabhu} and references therein).\\
\text { } \text { } There is a growing literature dealing with portfolio optimization problems in markets with nonconstant coefficients.
Most of these papers assume that the external process is a diffusion process itself, like in the established volatility model of Heston \cite{Heston} or in the Ho-Lee and the Vasicek model of Korn and Kraft \cite{KKr}. B\"auerle and Rieder \cite{BRied} and Rieder and B\"auerle \cite{RBa} studied the portfolio optimization problem with an observable and an unobservable Markov-modulated drift, respectively. This problem under stochastic volatility was considered by Pham and Quenez \cite{PQu} and by Fleming and Hern\'andez-Hern\'andez \cite{FHer}. In contrast to diffusion volatility, Markov chain volatility has the advantage that many portfolio problems can be solved explicitly. Moreover, a diffusion process can be approximated arbitrarily closely by a continuous-time Markov chain (see Kushner and Dupuis \cite{KDup}).\\
\text { } \text { } Portfolio optimization problems have also been studied in financial markets with regime switching.
One of the first papers is Zariphopoulou \cite{Zar} in which the author maximizes the utility of consumption under proportional transaction costs in a market where stock returns are determined by a continuous-time Markov chain, and establishes a viscosity property of the value function.
The results of Zariphopoulou were extended by many authors, among them B\"auerle and Rieder \cite{BRied}, Zhang and Yin \cite{ZYin} and Stockbridge \cite{Sto}.
To solve the problem of maximizing the investor's expected utility of terminal wealth, some authors use numerical methods (see Sass and Haussmann \cite{SHa}, Nagai and Runggaldier \cite{NRun}, Shen and Siu \cite{SSiu}, Fu et al. \cite{FWYang}).
Several authors have combined regime switching with different types of frictions. Gassiat et al. \cite{GGPh} analyzed a utility maximization problem in
a Black-Scholes market with regime switching under liquidity constraints. Jang et al. \cite{JKKLL} investigated the portfolio selection with transaction costs. They investigated the impact of the interaction of these two effects on the optimal portfolio and found that the impact of transaction costs on the optimal portfolio becomes more pronounced in the presence of switching regimes.
In turn Zhang et al. \cite{ZSMe} solved the portfolio selection problem without transition cost in a continuous-time Markovian regime switching Black-Scholes-Merton market. They obtained closed-form solutions for the optimal portfolio strategies in the cases of the logarithmic utility and the power utility. Similar results for a Black-Scholes market with regime switching were obtained by Liu \cite{Liu}, Guo et al. \cite{GuoMM} and Sotomayor and Cadenillas \cite{SCad}.
A discrete time set up was also considered by Yin and Zhou \cite{YinZhou}. For the mean-variance portfolio selection problem of this type we refer to Zhou and Yin \cite{ZhouYin}.
Regime switching was also analyzed by Tu \cite{Tu} in a Bayesian setting with model uncertainty and parameter uncertainty. He showed that the economic cost of ignoring regime switching can exceed 2 percent per year. Bae et al. \cite{BKMu} constructed a program to optimize portfolios in the above mentioned framework and used it to show that the regime information helps avoid risk during left-tail events.
\vb

\text { } \text { } The goal of this paper is to construct a general approach of building the optimal portfolio taking into account the asset jumps and possibility of changing environment
by considering asset prices modelled by It\^{o}-Markov additive processes.
In particular, we assume that the interest rate and the volatility of the financial assets depend on a continuous-time finite-state Markov chain.
Thus our model takes into account two sources of risk: the jump diffusion risk and the regime switching risk.
The jump diffusion risk refers to the source of risk due to fluctuations of market prices modelled by a Poisson random measure,
while the regime switching risk refers to the source of risk due to transitions of economic conditions.\\
\text { } \text { } Due to the presence of these sources of risk our market model is incomplete.
In this paper we show how to complete the It\^{o}-Markov additive market
model by adding Markovian jump securities, Markovian power-jump securities and impulse regime switching securities.
Using these securities all contingent claims can be replicated by a self-financing portfolio.
The main idea of completing a Markovian regime switching market is inspired by Corcuera et al. \cite{CGNS, CNS}, Guo \cite{Guo}, Karatzas et al. \cite{KLSXu}, Niu \cite{Niu} and Zhang et al. \cite{ZESGuo}.
 Moreover, we give conditions for the market to be asymptotic-arbitrage-free, namely, we find a martingale measure under which all the discounted price processes are martingales.\\
\text { } \text { } In this paper we also consider the problem of identifying
the optimal strategy that maximizes the expected value of the utility function of the wealth process at the end of some fixed period.
The analysis is conducted for the logarithmic and power utility functions.
To solve the main problem of determining the optimal portfolio we do not use dynamic programming but the direct differentiation approach.\\
\text { } \text { } This paper is organized as follows. In Section 2, we present the dynamics of the price process in an It\^{o}-Markov additive market. In Section 3, we enlarge this market by Markovian jump securities, Markovian power-jump securities and impulse regime switching securities. In Sections 4 and 5, we show that the enlarged market is asymptotic-arbitrage-free and complete. In Section 6 we state the portfolio optimization problem and solve it for the power utility and the logarithmic utility function.
Moreover, Section 6 gives a relationship between finite and infinite markets.

\section{Market model }\label{prel}
\text { } \text { } Let $(\Omega, \mathcal{F}, \mathbb{P})$ be a complete
probability space and let $\mathbb{T} := [0,T]$, for fixed $0<T<\infty$, represents the maturity time for all economic activities. On this probability space we consider the observable and continuous-time Markov chain  $J :=\{J(t): t\in \mathbb{T}\} $ with a finite state space. The role of the Markov chain is to ensure that the parameters change according to the market environment and the different states of the Markov chain represent the different states of the economy. For simplicity, we follow the notation of Elliott et al. \cite{EAM} and we identify the state space with the standard basis $E := \{\textbf {e}_{1}, \ldots , \textbf {e}_{N}\}$. Here $\textbf {e}_{i}\in\mathbb{R}^{N}$ and the $j$th component of $\textbf{e}_{i}$ is the Kronecker delta $\delta_{ij}$ for each $i,j = 1,  \ldots ,N$. Moreover, the Markov chain  $J$ is characterized by an intensity matrix $[\lambda_{ij}]_{i,j=1}^{N}$. The element $\lambda_{ij} $ is the transition intensity of the Markov chain $J$ jumping from state $\textbf {e}_{i}$ to state $\textbf {e}_{j}$.
We assume $\lambda_{ij}>0$ for $i\neq j$. Note that $\sum\limits_{j=1}^{N}\lambda_{ij}=0$, thus $\lambda_{ii}<0$.

\subsection{Risk-free asset}

Now we describe the dynamic of the price process of risk-free asset $B$ as follows:
\begin{equation} \label{145}
\mathrm{d}B(t)=r(t)B(t)\mathrm{d}t,\textrm{ } \textrm{ }\textrm{ }\textrm{ }\textrm{ } B(0)=1.
\end{equation}
Here $r$ is the interest rate of $B$ and it is modulated by Markov chain $J$
\begin{equation*}\label{1212}
r(t):= \langle  \textbf{r}, J(t) \rangle  = \sum_{i=1}^{N} r_{i} \langle \textbf{e}_{i},J(t) \rangle  ,
\end{equation*}
where $\textbf {r}=(r_{1}, \ldots ,r_{N})'\in\mathbb{R}^{N}_{+}$ and $\langle \cdot, \cdot\rangle$ is a scalar product in $\mathbb{R}^{N}$. The value $r_{i}>1$ represents the value of the interest rate when the Markov chain
is in the state space $\textbf{e}_{i}$.

\subsection{Risky asset}
We consider the market with a price of risky asset described by It\^{o}-Markov additive processes.\\
\text { } \text { } A process $(J,X)=\{(J(t),X(t)): t\in \mathbb{T}\}$ on the state space $\{\textbf{e}_1, \ldots , \textbf{e}_N\} \times \mathbb{R}$ is a {\it Markov additive process} (MAP) if $(J,X)$ is a Markov process and the conditional distribution of $(J(s+t),X(s+t)-X(s))$ for $s, t\in \mathbb{T}$, given $(J(s),X(s))$, depends only on $J(s)$ (see \c Cinlar \cite{Ca, Cb}).
Every MAP has a very special structure. It is usually said that $X$ is the additive component and $J$ is the background process representing the environment.
Moreover, the process $X$ evolves as a L\'{e}vy process while $J(t) = \textbf{e}_j$.\\
 \text { } \text { } Following Asmussen and Kella \cite{AK} we can decompose the process $X$ as follows:
\begin{equation}\label{defX}
X(t)=\overline{X}(t)+\overline{\overline{X}}(t),
\end{equation}
where
\begin{equation*}\label{ovvX}
\overline{\overline{X}}(t):=\sum_{i=1}^{N}\Psi_{i}(t)
\end{equation*}
for
\begin{equation}\label{Psij}
\Psi_{i}(t):=\sum_{n\geq1}U^{(i)}_{n}\textbf{1}_{\{J(T_{n})=\textbf {e}_{i},\textrm{ } T_{n}\leq t\}}
\end{equation}
and for the jump epochs $\{T_n\}$ of $J$.
Here $U^{(i)}_{n}$ $(n\geq 1, 1\leq i \leq N)$ are independent random variables which are also independent of $\overline{X}$ such that for every fixed $i$, the random variables $U^{(i)}_{n}$ are identically distributed.
Note that we can express the process $\Psi_{i}$ as follows:
\begin{equation*}
\Psi_{i}(t)=\int_0^t  \int_{\mathbb{R}}x \textrm{ } \Pi_{U}^{i} (\mathrm{d}s, \mathrm{d}x)
\end{equation*}
for the point measure
\begin{equation}\label{PiU}
\Pi_{U}^{i} ([0,t], \mathrm{d}x):=\sum_{n\geq1}  \mathbb{P}(U^{(i)}_{n} \in \mathrm{d}x )  \textbf{1}_{\{J(T_{n})=\textbf {e}_{i},\textrm{ } T_{n}\leq t\}}, \textrm{ }\textrm{ } i = 1, \ldots ,N.
\end{equation}
Moreover, we define the compensated point measure $\bar{ \Pi}_{U}^{i} (\mathrm{d}t, \mathrm{d}x) := \Pi_{U}^{i}(\mathrm{d}t, \mathrm{d}x) - \lambda_{i}(t)\eta_{i}(\mathrm{d}x)\mathrm{d}t  $  for $\lambda_{j}(t):=\sum\limits_{i\neq j} \textbf{1}_{\{J(t-)=  \textbf{e}_{i}\}} \lambda_{ij}$  and $\eta_{i}(\mathrm{d}x)= \mathbb{P}(U^{(i)}_{n} \in \mathrm{d}x ) $.

\begin{remark}
One can consider
jumps $U^{(ij)}$ with distribution depending also on the state $\textbf{e}_j$ the Markov chain is jumping to by
extending the state space to the pairs $(\textbf{e}_i, \textbf{e}_j)$ (see Gautam et al. \cite[Thm. 5]{Palmowskietal}
for details).
\end{remark}
\text { } \text { } The first component in definition (\ref{defX}) is an It\^{o}-L\'evy process and it has the following decomposition (see Oksendal and Sulem \cite[p. 5]{OS}):
\begin{equation} \label{1010}
\overline{X}(t):= \overline{X}(0)+ \int_{0}^{t} \mu_{0}(s)\mathrm{d}s + \int_{0}^{t}\sigma_{0}(s)\mathrm{d}W(s)+ \int_{0}^{t}  \int_{\mathbb{R}} \gamma(s-,x) \bar{\Pi}(\mathrm{d}s,\mathrm{d}x),
\end{equation}
where $W$ denotes the standard Brownian motion independent of $J$ and $ \bar{\Pi}(\mathrm{d}t,\mathrm{d}x):=\Pi(\mathrm{d}t,\mathrm{d}x)-\nu(\mathrm{d}x)\mathrm{d}t$ is the compensated Poisson random measure which is independent of $J$ and $W$. Furthermore, we define
\begin{equation}\label{m0}
 \mu_{0}(t):=\langle  \boldsymbol\mu_{0}, J(t) \rangle=\sum_{i=1}^{N}\mu_{0}^{i}\langle \textbf{e}_{i},J(t) \rangle ,
 \end{equation}
\begin{equation}\label{sigma0}
\sigma_{0}(t):=\langle \boldsymbol\sigma_{0}, J(t) \rangle =\sum_{i=1}^{N}\sigma_{0}^{i} \langle \textbf{e}_{i},J(t) \rangle ,
\end{equation}
\begin{equation}\label{gamma0}
\gamma(t,x):=\langle  \boldsymbol\gamma(x), J(t) \rangle=\sum_{i=1}^{N}\gamma_{i}(x) \langle \textbf{e}_{i}, J(t) \rangle
\end{equation}
for some vectors
$\boldsymbol\mu_{0}:=(\mu_{0}^{1}, \ldots ,\mu_{0}^{N})'\in \mathbb{R}^{N}$, $\boldsymbol\sigma_{0}:=(\sigma_{0}^{1}, \ldots , \sigma_{0}^{N})'\in\mathbb{R}^{N}_{+}$ and the vector-valued measurable function $\boldsymbol\gamma(x):=\big(\gamma_{1}(x), \ldots ,\gamma_{N}(x)\big)$.
The measure $\nu$ is the so-called jump-measure identifying the distribution of the sizes of the jumps of the Poisson measure $\Pi$. The components $\overline{X}$ and $\overline{\overline{X}}$ in \eqref{defX} are conditionally, on the state of the Markov chain $J$, independent. \\
\text { } \text { } Additionally, we suppose that the L\'{e}vy measure satisfies, for some $ \varepsilon >0$ and $\varrho >0$,
\begin{equation}\label{42}
\int_{(- \varepsilon,  \varepsilon)^{c}} \exp\big(\varrho |\gamma(s-,x)|\big)\nu(\mathrm{d}x)<\infty,\quad \int_{(- \varepsilon,  \varepsilon)^{c}} \exp(\varrho x)\mathbb{P}\big(U^{(i)}\in \mathrm{d}x\big)<\infty,
\end{equation}
for $i=1, \ldots,N$.
This implies that
\begin{equation*}
 \int_{\mathbb{R}}|\gamma(s-,x)|^{k}\nu(\mathrm{d}x)<\infty,\quad \mathbb{E}\big(U^{(i)}\big)^k<\infty, \textrm{ }\textrm{ } \textrm{ }\textrm{ }i=1, \ldots,N, \textrm{ }\textrm{ }  k \geq 2,
\end{equation*}
and that the characteristic function $\mathbb{E}[\exp(ku X)]$ is analytic in a neighbourhood of $0$. Moreover, $X$
has moments of all orders and the polynomials are dense in $L^{2}(\mathbb{R}, \mathrm{d}\varphi (t,x))$, where $\varphi(t,x):=\mathbb{P}\big(X(t)\leq x\big)$.\\
\text { } \text { } Now we are ready to define the main process in this paper.
The process  $(J,X)= \{(J(t),X(t)):t \in \mathbb{T} \}$ (for simplicity sometimes we write only $X$) with the decomposition (\ref{defX}) is called an {\it It\^{o}-Markov additive process}.\\
\text { } \text { } This process evolves as the It\^{o}-L\'evy process $\overline{X}$ between changes of states of the Markov chain $J$, that is, its parameters depend on the current state $\textbf{e}_i$ of the Markov chain $J$.
In addition, transition of $J$ from $\textbf{e}_i$ to $\textbf{e}_j$ triggers a jump of $X$ distributed as $U^{(i)}_n$. This is a so-called non-anticipative It\^{o}-Markov additive process.\\
\text { } \text { } It\^{o}-Markov additive processes are a natural generalization of It\^{o}-L\'evy processes and thus of L\'evy processes.
Moreover, the structure \eqref{defX}
explains the used name and can be seen
as {\it Markov-modulated It\^{o}-L\'evy process}.
Indeed, if $\gamma(s,x)=x$ then $X$ is a Markov additive process. If additionally $N=1$, then $X$ is a L\'evy process.
If $U^{(i)}\equiv 0$ and $N> 1$ then $X$ is a Markov modulated L\'{e}vy process (see Pacheco et al. \cite{PTP}).
If there are no jumps, that is, $ \bar{\Pi}(\mathrm{d}s, \mathrm{d}x)=0$, we have a Markov modulated Brownian motion.
\vb

\text { } \text { } To describe the price of the risky asset we use the It\^{o}-Markov additive process.
We interpret the coefficient  $\mu_{0} $ defined in (\ref{m0}) as the appreciation rate and $\sigma_{0} $ defined in (\ref{sigma0}) as the volatility of the risky asset for each $i=1, \ldots ,N$.
 In a similar way, $\mu_{0}^{i}$ and $\sigma_{0}^{i}$ represent the appreciation rate and the volatility of the risky asset , respectively, when the Markov chain is in state $\textbf e_{i}$. The condition
 \begin{equation*}\label{1372}
\mu_{0}^{i}>r_{i}, \textrm{ }\textrm{ } \textrm{ }\textrm{ }i=1,\ldots,N,
\end{equation*}
is required to avoid arbitrage opportunities in the market.
 We assume the evolution of the price process of the risky asset $S_{0}$ is governed by the It\^{o}-Markov additive  process as follows:
\begin{equation}\label{555}
\begin {cases}
\mathrm{d}S_{0}(t)=S_{0}(t-)\bigg[ \mu_{0}(t)\mathrm{d}t+\sigma_{0}(t)\mathrm{d}W(t)+ \int_{\mathbb{R}}\gamma(t-,x) \bar{\Pi}(\mathrm{d}t,\mathrm{d}x)+ \sum\limits_{i=1}^{N}  \int_{\mathbb{R}}x \bar{\Pi}_{U}^{i} (\mathrm{d}t, \mathrm{d}x)\bigg],\\
S_{0}(0)=s_{0}>0.\\
\end{cases}
\end{equation}
\section{Enlarging the It\^{o}-Markov additive market }
\text { } \text { } Now we enlarge the primary market by some financial assets: Markovian jump securities, Markovian power-jump securities and impulse regime switching securities in order to complete the market.\\
From now, we will work with the following filtration on $(\Omega, \mathcal{F}, \mathbb{P})$:
\begin{equation*}
\mathcal{F}_{t}:= \mathcal{G}_{t} \vee  \mathcal{N},
\end{equation*}
where $ \mathcal{N}$ are the $\mathbb{P}$-null sets of $\mathcal{F}$ and
\begin{equation*}
\mathcal{G}_{t}:=\sigma\{J(s),W(s), \Gamma(s), \Pi_{U}^{1} ([0,s], \mathrm{d}x), \ldots, \Pi_{U}^{N} ([0,s], \mathrm{d}x) ;s\leq t\}
\end{equation*}
for
\begin{equation*}
\Gamma(t):= \int_{0}^{t}  \int_{\mathbb{R}} \gamma(s-,x) \Pi(\mathrm{d}s,\mathrm{d}x).
\end{equation*}
Note that the filtration $ \{\mathcal{F}_{t} \}_{t\geq 0}$ is right-continuous (see Karatzas and Shreve \cite[Prop. 7.7 ]{KS} and also Protter \cite[Thm. 31 ]{P}).
By the same arguments as in the proof of Thm. 3.3. of Liao \cite{Liao},
the filtration $\{ \mathcal{G}_{t}\}_{t\geq 0}$ is equivalent to
\begin{equation*}
\sigma\{J(s), \overline{X}(s), \Pi_{U}^{1} ([0,s], \mathrm{d}x), \ldots, \Pi_{U}^{N} ([0,s], \mathrm{d}x) ;s\leq t\}.
\end{equation*}
\subsection{Markovian jump securities}
Let ${T_{n}}$ $(n = 1, 2, \ldots )$ denote the jump epochs of the chain $J$, where $0 \leq T_{1} \leq T_{2} \leq \ldots $
We observe that the Markov chain $J$ can be represented in terms of a marked point process $\Phi_{j} $ defined by
\begin{equation*}
\Phi_{j}(t):=\Phi([0,t] \times \textbf e_{j})=\sum_{n \geq 1}\textbf{1}_{\{J(T_{n}) =  \textbf {e}_{j}, \textrm{ }   T_{n} \leq t \}}, \textrm{ }\textrm{ } j = 1, \ldots ,N.
\end{equation*}
Note that the process $\Phi_{j} $ describes the number of jumps into state $\textbf{e}_{j}$ up to time $t$. Let $\phi_{j}$ be the dual predictable projection of $\Phi_{j}$ (sometimes called the compensator). That is, the process
\begin{equation}\label{ovlPhi}
\overline{\Phi}_{j}(t):=\Phi_{j}(t)-\phi_{j}(t),\textrm{ }\textrm{ } j = 1, \ldots ,N,
\end{equation}
is a $\{\mathcal{F}_{t}\}$-martingale and it is called the $j$th Markovian jump martingale.
Note that $\phi_{j} $ is unique and
\begin{equation*}\label{9878}
\phi_{j}(t):= \int_{0}^{t}\lambda_{j}(s)\mathrm{d}s,
\end{equation*}
for
\begin{equation}\label{lamb}
\lambda_{j}(t):=\sum\limits_{i\neq j} \textbf{1}_{\{J(t-)=  \textbf{e}_{i}\}} \lambda_{ij}
\end{equation}
(see Zhang et al. \cite[p. 290]{ZESGuo}).\\
Now we consider geometric Markovian jump securities $S_{j}$ (for j = 1, \ldots, N) with evolution of prices described by marked point martingales as follows:
\begin{equation}\label{45}
\begin {cases}
\mathrm{d}S_{j}(t)=S_{j}(t-)\Big[\mu_{j}(t)\mathrm{d}t+\sigma_{j}(t-)\mathrm{d}\overline{\Phi} _{j}(t)\Big],\\
S_{j}(0)>0,\\
\end{cases}
\end{equation}
where the appreciation rate $\mu_{j}$ and the volatility $\sigma_{j}$ are given by
$$ \mu_{j}(t):=\langle  \boldsymbol\mu_{j}, J(t) \rangle=\sum_{i=1}^{N}\mu_{j}^{i} \langle \textbf{e}_{i},J(t) \rangle , $$
$$\sigma_{j}(t):=\langle \boldsymbol\sigma_{j}, J(t) \rangle =\sum_{i=1}^{N}\sigma_{j}^{i} \langle \textbf{e}_{i},J(t) \rangle  $$
with $\boldsymbol\mu_{j}:=(\mu_{j}^{1}, \ldots ,\mu_{j}^{N})'\in \mathbb{R}^{N}$ and  $\boldsymbol\sigma_{j}:=(\sigma_{j}^{1},  \ldots ,\sigma_{j}^{N})'\in\mathbb{R}_{+}^{N}$.

\subsection{Markovian power-jump securities}

Following Corcuera et al. \cite{CNS} we introduce the power-jump processes
$$X^{(k)}(t):= \sum\limits_{0<s\leq t}(\Delta \overline{X}(s))^{k},  \textrm{ } \textrm{ } \textrm{ }\textrm{ }\textrm{ }\textrm{ } k \geq 2,$$
where $\Delta \overline{X}(s)=\overline{X}(s)-\overline{X}(s-)$. We set $X^{(1)}(t)=\overline{X}(t)$.
The process $X^{(k)}$ is also an It\^{o}-L\'evy process with the same jump times as the original process $\overline{X}$ but with their sizes being
the $k$th powers of the jump sizes of $\overline{X}$. From Protter \cite[p. 29]{P} we have
\begin{equation*}
\mathbb{E}\big[X^{(k)}(t)\big|  \mathcal{J}_{t} \big]=\mathbb{E}\bigg( \sum_{0<s\leq t} (\Delta \overline{X}(s))^{k} \big|  \mathcal{J}_{t}  \bigg)=
\int_0^t \int_{\mathbb{R}}\gamma^{k}(s-,x)\nu(\mathrm{d}x)\mathrm{d}s <\infty, \mathbb{P}-a.e. \textrm{ } \textrm{ } \textrm{ }\textrm{ }\textrm{ }\textrm{ } k \geq 2,
\end{equation*}
for $\mathcal{J}_{t}:=\sigma\{ J(s): s \leq t\}$. Hence the processes
\begin{equation*}
\overline{X}^{(k)}(t):=X^{(k)}(t)-\int_0^t  \int_{\mathbb{R}}\gamma^{k}(s-,x)\nu(\mathrm{d}x)\mathrm{d}s, \textrm{ } \textrm{ } \textrm{ }\textrm{ }\textrm{ }\textrm{ } k \geq 2,
\end{equation*}
are $\{\mathcal{F}_{t}\}$-martingales (called Teugels martingales of order $k$; see Schoutens \cite{S} for details).
Indeed, since $\overline{X}$ is $\{\mathcal{F}_{t}\}$-adapted by Jacod and Shiryaev \cite[Prop. 1.25]{JS} the process $\Delta \overline{X}$ is also
$\{\mathcal{F}_{t}\}$-adapted. Furthermore, the integral of an $\{\mathcal{F}_{t}\}$-adapted process with respect to an $\{\mathcal{F}_{t}\}$-adapted stochastic measure or an $\{\mathcal{F}_{t}\}$-adapted stochastic process is still $\{\mathcal{F}_{t}\}$-adapted (see Jacod and Shiryaev \cite[Prop. 3.5 and Thm. 4.31(i)]{JS}). Hence $X^{(k)}$ is $\overline{X}^{(k)}$ are $\{\mathcal{F}_{t}\}$-adapted for $k \geq 2$.

\text { } \text { } We additionally enlarge the market with a series of Markovian $k$th-power-jump assets \linebreak $S^{(k)}$ (for $k \geq 2$).
The price process of $S^{(k)}$ is described by the stochastic differential equation
\begin{equation}\label{46}
\begin {cases}
\mathrm{d}S^{(k)}(t)=S^{(k)}(t-)\Big[\mu^{(k)}(t)\mathrm{d}t+\sigma^{(k)}(t-)\mathrm{d}\overline{X}^{(k)}(t)\Big],\\
S^{(k)}(0)>0,\\
\end{cases}
\end{equation}
where the coefficients are determined by the Markov chain $J$, namely:
$$ \mu^{(k)}(t):=\langle  \boldsymbol\mu^{(k)}, J(t) \rangle=\sum_{j=1}^{N}\mu_{j}^{(k)} \langle \textbf{e}_{j},J(t) \rangle \text{    } \text{    } \text{    }  \text{ and   } \text{    } \text{    } \text{    } \sigma^{(k)}(t):=\langle \boldsymbol\sigma^{(k)}, J(t) \rangle =\sum_{j=1}^{N}\sigma_{j}^{(k)} \langle \textbf{e}_{j},J(t) \rangle  $$
for $\boldsymbol\mu^{(k)}:=(\mu_{1}^{(k)}, \ldots ,\mu_{N}^{(k)})'\in \mathbb{R}^{N}$ and $\boldsymbol\sigma^{(k)}:=(\sigma_{1}^{(k)},  \ldots ,\sigma_{N}^{(k)})'\in\mathbb{R}_{+}^{N}$.

\subsection{Impulse regime switching securities}
We will also need power martingales related to the second component of $X$ given in \eqref{defX}, namely to $\overline{\overline{X}}$ or to $\Psi_i$, defined in \eqref{Psij}. For $l \geq 1$ and $i=1, \ldots , N$ we define
\begin{equation*}\label{Psijk}
\Psi_{i}^{(l)}(t): =\sum_{n\geq1}\left(U^{(i)}_{n}\right)^l\textbf{1}_{\{J(T_{n})=\textbf {e}_{i},\textrm{ } T_{n}\leq t\}}=\int_0^t  \int_{\mathbb{R}}x^{l} \textrm{ } \Pi_{U}^{i} (\mathrm{d}s, \mathrm{d}x)
\end{equation*}
for $\Pi_{U}^{i}$ given by (\ref{PiU}).
The compensated version of $\Psi_{i}^{(l)}$ is called an impulse regime switching martingale if
\begin{equation*}\label{90}
\overline{\Psi}_{i}^{(l)}(t):= \Psi_{i}^{(l)}(t) - \mathbb{E}\big(U^{(i)}_{n}\big)^{l}\phi_{i}(t)=\int_0^t \int_{\mathbb{R}}x^{l} \textrm{ }\bar{ \Pi}_{U}^{i} (\mathrm{d}s, \mathrm{d}x),
\end{equation*}
where $\bar{ \Pi}_{U}^{i} (\mathrm{d}t, \mathrm{d}x) = \Pi_{U}^{i}(\mathrm{d}t, \mathrm{d}x) - \lambda_{i}(t)\eta_{i}(\mathrm{d}x)\mathrm{d}t  $  for $\lambda_{i}$ defined in (\ref{lamb}) and $\eta_{i}(\mathrm{d}x)= \mathbb{P}(U^{(i)}_{n} \in \mathrm{d}x ) $.
Using similar arguments to those above it follows that $\overline{\Psi}_{i}^{(l)}$ is an $\{\mathcal{F}_{t}\}$-martingale for $l \geq 1$ and $i=1, \ldots ,N$.\\

\text { } \text { } We characterize the evolution of impulse regime switching securities $S_{i}^{(l)}$ as follows:
\begin{equation}\label{477}
\begin {cases}
\mathrm{d}S_{i}^{(l)}(t)=S_{i}^{(l)}(t-)\Big[\mu_{i}^{(l)}(t)\mathrm{d}t+\sigma_{i}^{(l)}(t-)\mathrm{d}\overline{\Psi}_{i}^{(l)}(t)\Big],\\
S_{i}^{(l)}(0)>0,\\
\end{cases}
\end{equation}
where the coefficients are determined by the Markov chain $J$, namely:
$$ \mu_{i}^{(l)}(t):=\langle  \boldsymbol\mu_{i}^{(l)}, J(t) \rangle=\sum_{j=1}^{N}\mu_{i,j}^{(l)} \langle \textbf{e}_{j},J(t) \rangle; \text{    } \text{    } \text{    }  \text{    } \text{    } \sigma_{i}^{(l)}(t):=\langle \boldsymbol\sigma_{i}^{(l)}, J(t) \rangle =\sum_{j=1}^{N}\sigma_{i,j}^{(l)} \langle \textbf{e}_{j},J(t) \rangle$$
for $\boldsymbol\mu_{i}^{(l)}:=(\mu_{i,1}^{(l)}, \ldots  ,\mu_{i,N}^{(l)})'\in \mathbb{R}^{N}$ and  $\boldsymbol\sigma_{i}^{(l)}:=(\sigma_{i,1}^{(l)}, \ldots  ,\sigma_{i,N}^{(l)})'\in\mathbb{R}^{N}_{+}$  $(i=1, \ldots  ,N$ and $l \geq 1)$.

\text { } \text { } Combining (\ref{145}), (\ref{555}), (\ref{45}), (\ref{46}) and (\ref{477}) we get an enlarged It\^{o}-Markov additive market:
\begin{equation}\label{47}
\begin {cases}
\mathrm{d}B(t)=r(t)B(t)\mathrm{d}t, \\
\mathrm{d}S_{0}(t)=S_{0}(t-)\bigg[\mu_{0}(t)\mathrm{d}t+\sigma_{0}(t)\mathrm{d}W(t)+ \int_{\mathbb{R}} \gamma (t-,x)\bar{\Pi}(\mathrm{d}t,\mathrm{d}x)+ \sum\limits_{i=1}^{N}  \int_{\mathbb{R}}x \bar{\Pi}_{U}^{i} (\mathrm{d}t, \mathrm{d}x) \bigg],\\
\mathrm{d}S_{j}(t)=S_{j}(t-)\Big[\mu_{j}(t)\mathrm{d}t+\sigma_{j}(t-)\mathrm{d}\overline{\Phi} _{j}(t)\Big],\\
\mathrm{d}S^{(k)}(t)=S^{(k)}(t-)\Big[\mu^{(k)} (t)\mathrm{d}t+\sigma^{(k)}(t-)\mathrm{d}\overline{X}^{(k)}(t)\Big],\\
\mathrm{d}S_{i}^{(l)}(t)=S_{i}^{(l)}(t-)\Big[\mu_{i}^{(l)} (t)\mathrm{d}t+\sigma_{i}^{(l)}(t-)\mathrm{d}\overline{\Psi}_{i}^{(l)}(t)\Big],
\end{cases}
\end{equation}
for $i,j=1, \ldots ,N $, $k \geq 2$ and $l \geq 1$. In Section \ref{009}, we will prove that under a new martingale measure this market is complete. Note that $\mu_{j}$, $\mu^{(k)}$, $\mu_{i}^{(l)}$,  $\sigma_{j}$, $\sigma^{(k)} $ and $\sigma_{i}^{(l)}$ are  artificial parameters and can be changed later.
\vb

\text { } \text { } Corcuera et al. \cite{CNS} motivate trading in power-jump assets as follows. A power-jump process of order two is just a variation process of degree two, i.e. a quadratic variation process (see Barndorff-Nielsen and Shephard \cite{BNSh, BNSh2}), and is related to the so-called realized variance. Contracts on realized variance have found their way into OTC markets and are now traded regularly. Typically a 3th-power-jump asset measures a kind of asymmetry ("skewness") and a 4th-power-jump process measures extremal movements ("kurtosis"). Trade in such assets can be of use if one likes to bet on the realized skewness or realized kurtosis of the stock. Furthermore, an insurance contract against a crash can also be easily built from 4th-power-jump (or $i$th-power-jump, $i > 4$) assets.
One can also consider financial-insurance contracts that hedge L\'evy jumps, e.g. COS and CDS contracts.

\section{Martingale measure and asymptotic arbitrage }
$\textrm{ } \textrm{ }$ Considering a financial market containing an infinite number of assets, Kabanov and Kramkov \cite{KK1} introduced the notion of large financial market. This type of market is described by a sequence of market models with a finite number of securities each, also called small markets. In \cite{KK1}, the authors introduce an extension of the classical approach to arbitrage theory, namely arbitrage in a large financial market, called asymptotic arbitrage.
A deep study of asymptotic arbitrage has been carried out by Kabanov and Kramkov \cite{KKK} and Bj\"ork and N\"aslund \cite{BN}.\\
$\textrm{ } \textrm{ }$ In this section we identify a martingale measure in our It\^{o}-Markov additive market and prove that this market model is asymptotic-arbitrage-free.\\
$\textrm{ } \textrm{ }$ Let us start with the definition of {\it asymptotic arbitrage}.
An asymptotic arbitrage is when we have a sequence of strategies such that, for some real number $c > 0$, the value process $V^{n}$ on a finite market satisfies:
\begin{itemize}
\item $V^n(t) \geq -c$ for each $0<t \leq T$ and for each $ n \in  \mathbb{N} $,
\item $V^n(0)=0$ for each $ n \in  \mathbb{N} $,
\item $\liminf\limits_{n \rightarrow \infty} V^n(T) \geq 0$, $\mathbb{P}$-a.s,
\item $\mathbb{P}\big(\liminf\limits_{n \rightarrow \infty} V^n(T) > 0 \big)>0$.
\end{itemize}

\begin{proposition} (Bj\"ork and N\"aslund \cite[ Prop. 6.1]{BN}). \label{22233}
If there exists a martingale measure $\mathbb{Q}$ equivalent to $\mathbb{P}$ then the market is asymptotic-arbitrage-free.
\end{proposition}

\text { } \text { }  Now we will find a measure $\mathbb{Q}$ under which the discounted price processes are martingales.\\
\text { } \text { }  Let $\mathcal{L}^{2}(W)$ be the set of all predictable, $\{\mathcal{F}_{t}\}$-adapted processes $\xi$ such that $\mathbb{E} \int_{0}^{T} \xi^{2}(s) \mathrm{d}s< \infty $. In a similar way we define $\mathcal{L}^{1}(\phi _{j} )$, that is, $\xi \in \mathcal{L}^{1}(\phi _{j} )$ iff $\xi$ is predictable, $\{\mathcal{F}_{t}\}$-adapted and satisfies $\mathbb{E} \int_{0}^{T} |\xi(s) |\lambda^{2}_{j} \mathrm{d}s< \infty $.
\begin{proposition} (Boel and Kohlmann \cite[p. 515]{BK}) \label{223}
Let $\psi_{0} \in \mathcal{L}^{2}(W)$
and $\psi_{j} \in \mathcal{L}^{1}(\phi _{j} ) $ for all $j=1, \ldots ,N$. Then
\begin{equation} \label{5}
\ell(t):=\exp \Bigg[ \int_{0}^{t} \psi_{0}(s)\mathrm{d}W(s) -\frac{1}{2}\int_{0}^{t} \psi_{0}^{2}(s)\mathrm{d}s-   \sum_{j=1}^{N}\int_{0}^{t} \psi_{j}(s)\phi _{j}(\mathrm{d}s) \Bigg]
\end{equation}
\begin{equation*}
 \times \prod_{j=1}^{N} \prod_{\substack{J(t-)\neq J(t)\\ J(t)=\textbf {e}_{j} } } (1+\psi_{j}(t))
\end{equation*}
is a local martingale. If additionally $\mathbb{E} \ell(t)=1 $ then it is a true martingale.
\end{proposition}

From now on, we assume that $\mathbb{E } \ell(t)=1 $.
Let $\mathbb{Q}$ be the probability measure defined by the Radon-Nikodym derivative
\begin{equation*} \label{6}
\ell(t)=\dfrac{\mathrm{d}\mathbb{Q}}{\mathrm{d}\mathbb{P}} \bigg\arrowvert _{\mathcal{F}_{t}}.
\end{equation*}
Then $\ell$, given in (\ref{5}), is the density process for the new  martingale measure $\mathbb{Q}$. By adding a superscript $\mathbb{Q}$ we denote processes observed under this new measure.
By a generalized version of Girsanov's theorem for jump-diffusion processes we have the following theorem:
\begin{theorem} (Boel and Kohlmann \cite[p. 517]{BK}) \label{100}
The process $\overline{X}$ given in (\ref{1010}) under the new martingale measure $\mathbb{Q}$ has the form
$$\overline{X}^{\mathbb{Q}}(t)=\int_{0}^{t}\sigma_{0}(s)\mathrm{d}W^{\mathbb{Q}}(s)+ \int_{0}^{t}  \int_{\mathbb{R}} \gamma(s-,x) \bar{\Pi}(\mathrm{d}s,\mathrm{d}x),$$
where
$$W^{\mathbb{Q}}(t)=W(t)-\int_{0}^{t}\psi_{0}(s)\mathrm{d}s $$
is a standard $\mathbb{Q}$-Brownian motion.\\
Moreover, for $j=1, \ldots ,N$, the process $\overline{\Phi}_{j}$ given in (\ref{ovlPhi}) under the measure $\mathbb{Q}$
is a martingale and takes the form
$$\overline{\Phi}_{j}^{\mathbb{Q}}(t)=\Phi_{j}(t)-\int_{0}^{t}\Big(1+\psi_{j}(s)\Big)\phi_{j}(\mathrm{d}s),$$
 that is, the unique predictable projection of $\Phi_{j}$ under $\mathbb{Q}$ is given by
$$\phi_{j}^{\mathbb{Q}}(t)=\int_{0}^{t}\Big(1+\psi_{j}(s)\Big)\phi_{j}(\mathrm{d}s). $$
\end{theorem}

\begin{remark}\label{rembo}
If $\psi_{0}$ and $\psi_{j}$ $(j=1, \ldots,N) $ are bounded, then $\mathbb{E} \ell(t)=1 $ (see the Novikov condition in Karatzas and Shreve \cite[Cor. 3.5.13, p. 199]{KS} and Resnick \cite[Thm. 5.1, p. 135]{Resn}).
\end{remark}

\text { } \text { } Note that $\overline{\Psi}_{i}^{(l)}$ for $i=1, \ldots ,N $,  $l \geq 1$, $\overline{X}^{(k)}$ for $k \geq 2$, $\bar{\Pi}_{U}^{i}$ and  $\bar{\Pi}$ do not change their
laws under the new measure $\mathbb{Q}$.
Moreover, under $\mathbb{Q}$ the price processes are represented as follows:
\begin{equation*}
\begin {cases}
\mathrm{d}B(t)=r(t)B(t)\mathrm{d}t, \\
\mathrm{d}S_{0}(t) = S_{0}(t-)\bigg[ \Big(\mu_{0}(t)+\sigma_{0}(t)\psi_{0}(t)\Big)\mathrm{d}t+\sigma_{0}(t)\mathrm{d}W^{\mathbb{Q}}(t)
 +  \int\limits_{\mathbb{R}}\gamma(t-,x) \bar{\Pi}(\mathrm{d}t,\mathrm{d}x)+ \sum\limits_{i=1}^{N}  \int\limits_{\mathbb{R}}x \bar{\Pi}_{U}^{i} (\mathrm{d}t, \mathrm{d}x)\bigg],\\
\mathrm{d}S_{j}(t) =  S_{j}(t-)\Big[\Big(\mu_{j}(t)+\sigma_{j}(t)\lambda_{j}(t)\psi_{j}(t)\Big)\mathrm{d}t+\sigma_{j}(t-) \mathrm{d}\overline{\Phi}_{j}^{\mathbb{Q}}(t)\Big],\\
\mathrm{d}S^{(k)}(t)=S^{(k)}(t-)\Big[\mu^{(k)} (t)\mathrm{d}t+\sigma^{(k)}(t-)\mathrm{d}\overline{X}^{(k)}(t)\Big],\\
\mathrm{d}S_{i}^{(l)}(t)=S_{i}^{(l)}(t-)\Big[\mu_{i}^{(l)} (t)\mathrm{d}t+\sigma_{i}^{(l)}(t-)\mathrm{d}\overline{\Psi}_{i}^{(l)}(t)\Big],
\end{cases}
\end{equation*}
for $i,j=1, \ldots ,N $, $k \geq 2$ and $l \geq 1$. Note that in the above equation we can take $r(t)=r(t-)$, $\mu_{j}(t)=\mu_{j}(t-)$ and $\sigma_{j}(t)=\sigma_{j}(t-)$ (for $j=0,1, \ldots ,N$).
In fact, by stochastic integration by parts, the discounted price processes are governed by
\begin{equation}\label{49}
\begin {cases}
\mathrm{d}\tilde{S}_{0}(t) = \tilde{S}_{0}(t-)\bigg[ \Big(\mu_{0}(t-)+\sigma_{0}(t-)\psi_{0}(t)-r(t-)\Big)\mathrm{d}t+\sigma_{0}(t)\mathrm{d}W^{\mathbb{Q}}(t)+  \int\limits_{\mathbb{R}}\gamma(t-,x) \bar{\Pi}(\mathrm{d}t,\mathrm{d}x)\\
\text{     } \text{     } \text{     } \text{     } \text{     }+ \sum\limits_{i=1}^{N}  \int\limits_{\mathbb{R}}x \bar{\Pi}_{U}^{i} (\mathrm{d}t, \mathrm{d}x) \bigg],\\
\mathrm{d}\tilde{S}_{j}(t) = \tilde{S}_{j}(t-)\Big[\Big(\mu_{j}(t-)+\sigma_{j}(t-)\lambda_{j}(t)\psi_{j}(t)-r(t-)\Big)\mathrm{d}t+\sigma_{j}(t-)\mathrm{d}\overline{\Phi}_{j}^{\mathbb{Q}}(t)\Big],\\
\mathrm{d}\tilde{S}^{(k)}(t)=\tilde{S}^{(k)}(t-)\Big[\Big(\mu^{(k)} (t)- r(t) \Big)\mathrm{d}t+\sigma^{(k)}(t-)\mathrm{d}\overline{X}^{(k)}(t)\Big],\\
\mathrm{d}\tilde{S}_{i}^{(l)}(t)=\tilde{S}_{i}^{(l)}(t-)\Big[\Big(\mu_{i}^{(l)} (t)- r(t) \Big)\mathrm{d}t + \sigma_{i}^{(l)}(t-)\mathrm{d}\overline{\Psi}_{i}^{(l)}(t)\Big],
\end{cases}
\end{equation}
where $\tilde{S}_{0}(t):=B^{-1}(t)S_{0}(t)$, $\tilde{S}_{j}(t):=B^{-1}(t)S_{j}(t)$, $\tilde{S}^{(k)}(t):=B^{-1}(t)S^{(k)}(t)$ and $\tilde{S}_{i}^{(l)}(t):=B^{-1}(t)S_{i}^{(l)}(t)$. Hence, we require $\tilde{S}_{0}$, $\tilde{S}_{j}$, $\tilde{S}^{(k)}$ and $\tilde{S}_{i}^{(l)}$ to be  martingales (for $i,j=1, \ldots ,N $, $k \geq 2$ and $l \geq 1$). A necessary and sufficient condition for this to hold is given by the following equations:
\begin{equation}\label{53}
\begin {cases}
\mu_{0}(t-)+\sigma_{0}(t-)\psi_{0}(t)-r(t-) = 0,\\
\mu_{j}(t-)+\sigma_{j}(t-)\lambda_{j}(t)\psi_{j}(t)-r(t-) = 0,\\
\mu^{(k)} (t)- r(t)=0,\\
\mu_{i}^{(l)} (t)- r(t)=0,
\end{cases}
\end{equation}
for $i,j=1, \ldots ,N $, $k \geq 2$ and $l \geq 1$.\\
\text { } \text { } Note that $\lambda_{j}(t)=0$ if $J(t-)=\textbf {e}_{j}$. Thus in this case, if $\mu_{j}^{j}\neq r_{j}$, the martingale condition would never be satisfied.
Therefore, the discounted price processes of all securities in the enlarged market would not be martingales under $\mathbb{Q}$. Thus we have to assume that $\mu_{j}^{j}=r_{j}$ for all $j=1, \ldots ,N $ to make the market asymptotic-arbitrage-free. From (\ref{53}), when $\lambda_{j}(t)\neq 0$ (i.e. $J(t-)\neq \textbf {e}_{j})$, the processes $\psi_{0}$ and $\psi_{j}$ are determined by
\begin{equation}\label{54}
\begin {cases}
\psi_{0}(t)  =  \dfrac{r(t-)-\mu_{0}(t-)}{\sigma_{0}(t-)},\\
\psi_{j}(t)  =  \dfrac{r(t-)-\mu_{j}(t-)}{\sigma_{j}(t-)\lambda_{j}(t)}, \textrm{ } \textrm{ } \textrm{ }\textrm{ } j=1, \ldots ,N.
\end{cases}
\end{equation}
\text { } \text { } Note that $\psi_{0}$ and $\psi_{j}$ $(j=1, \ldots,N) $ are bounded. Hence by Remark \ref {rembo} the density process $\ell$ is a true martingale.
Note that $\psi_{j}$ $(j=1, \ldots ,N )$ satisfies the assumptions of Proposition \ref{223}.
We can only determine $\psi_{j}$ when $J(t-)\neq \textbf {e}_{j}$ for $j= 1, \ldots ,N$ but this is sufficient to determine the equivalent martingale measure $\mathbb{Q}$. Indeed, if $J(t-)=\textbf {e}_{j}$ for $j= 1, \ldots ,N$, then $\phi _{j}(t)=0$ and $\psi_{j}$ has no influence on the value of the right side of (\ref{5}).
The above analysis yields the following theorem.
\begin{theorem}\label{29}
Assume that $\mu_{j}^{j}=r_{j}$ for all $j=1, \ldots ,N $ and $\psi_{0}$ and $\psi_{j}$ are given by (\ref{54}). Then the discounted price processes of the securities in the enlarged market (\ref{49}) are martingales under $\mathbb{Q}$ and this market is asymptotic-arbitrage-free.
\end{theorem}
\text { } \text { } From now on we assume that $\mu_{j}^{j}=r_{j}$ for all $j=1, \ldots ,N $.

\section{Asymptotic completeness of the enlarged market}\label{009}

\text { } \text { } Now we will analyze asymptotic completeness of the enlarged It\^{o}-Markov additive market.
A market is said to be complete if each claim can be replicated by a strategy, that is, the claim can be represented as a stochastic integral with respect to the asset prices.
We take as class of contingent claims the set $L^{2} (\Omega, \mathcal{F}, \mathbb{Q})$ of square integrable random variables under the equivalent martingale measure; then a self-financing strategy will be represented as an integrable process and the value of a self-financing portfolio will be represented as the stochastic integral of the strategy with respect to the assets.
In the case of market models with an infinite number of assets we define completeness in terms of approximate replication of claims. \\
\text { } \text { }  For finite market asset, completeness is equivalent to uniqueness of the equivalent martingale measure.
In the case of large markets this property does not occur.
Artzner and Heath \cite{AHeath} constructed a financial market with countably many securities for which there are two equivalent martingale measures under which the market is approximately complete.
In the context of a large financial market, B\"attig \cite{Batting} and B\"attig and Jarrow \cite{BJAr} proposed a definition of completeness which is independent of either the notion of arbitrage-free or equivalent martingale measures.  B\"attig \cite{Batting} also provided an example where the existence of an equivalent martingale measure can exclude the possibility of replicating a claim, to show how the two notions of arbitrage-free and completeness are independent in practice.

\text { } \text { } Under $\mathbb{Q}$ the price processes of the securities in the arbitrage-free market have the following representations:
\begin{equation}\label{55}
\begin {cases}
\mathrm{d}B(t) = r(t)B(t)\mathrm{d}t,\\
\mathrm{d}S_{0}(t) = S_{0}(t-)\bigg[r(t)\mathrm{d}t+\sigma_{0}(t)\mathrm{d}W^{\mathbb{Q}}(t)+ \int_{\mathbb{R}}\gamma(t-,x) \bar{\Pi}(\mathrm{d}t,\mathrm{d}x)+ \sum\limits_{i =1}^{N}  \int_{\mathbb{R}}x \bar{\Pi}_{U}^{i} (\mathrm{d}t, \mathrm{d}x) \bigg],\\
\mathrm{d}S_{j}(t) = S_{j}(t-)\Big[r(t)\mathrm{d}t+\sigma_{j}(t-)\mathrm{d}\overline{\Phi}_{j}^{\mathbb{Q}}(t)\Big],\\
\mathrm{d}S^{(k)}(t) = S^{(k)}(t-)\Big[r(t)\mathrm{d}t+\sigma^{(k)}(t-)\mathrm{d} \overline{X}^{(k)}(t)\Big],\\
\mathrm{d}S_{i}^{(l)}(t)=S_{i}^{(l)}(t-)\Big[r(t)\mathrm{d}t+\sigma_{i}^{(l)}(t-)\mathrm{d}\overline{\Psi}_{i}^{(l)}(t)\Big],
\end{cases}
\end{equation}
for $i,j=1, \ldots ,N $, $k \geq 2$ and $l \geq 1$.\\
\text { } \text { } We will show that the enlarged market (\ref{55}) is asymptotically complete in the sense that for every square-integrable contingent claim $A$ (i.e. a non-negative square-integrable random variable in $ L^{2} (\Omega, \mathcal{F}, \mathbb{Q})$) we can set up a sequence of self-financing portfolios whose final values converge in $ L^{2} (\Omega, \mathcal{F}, \mathbb{Q})$ to $A$.

\text { } \text { } These portfolios will consist of a finite number of risk-free asset, risky asset, $k$th-power-jump assets, $j$th geometric Markovian jump security and impulse regime switching securities.
We will make use of the following Martingale Representation Property.
\begin{theorem}(Palmowski, Stettner and Sulima \cite{PSS})\label{94}
Any square-integrable, $\{\mathcal{F}_{t}\}$-adapted $\mathbb{Q}$-martingale $M$ can be represented as follows:
\begin{eqnarray}\label{Theorem7}
M(t) &=& M(0)+ \int_{0}^{t} h_{0}(s)\mathrm{d}X^{\mathbb{Q}}(s)+ \sum_{j=1}^{N}\int_{0}^{t}h_{j}(s)\mathrm{d}\overline{\Phi}_{j}^{\mathbb{Q}}(s)+ \sum_{k=2}^{\infty}\int_{0}^{t}h^{(k)}(s)\mathrm{d}\overline{X}^{(k)}(s)\\
\nonumber & & +\sum_{i=1}^{N}\sum_{l=1}^{\infty}\int_{0}^{t} h^{(l)}_{i }(s)\mathrm{d}\overline{\Psi}_{i}^{(l)}(s),
\end{eqnarray}
where $h_{0}$, $h_{j}$, $h^{(k)}$ and $h^{(l)}_{i }$ are predictable processes (for $i,j=1, \ldots ,N $, $k \geq 2$ and $l \geq 1$).
\end{theorem}

\begin{remark}\label{rem34}\rm
The right-hand side of (\ref{Theorem7}) is understood as follows. We take finite sums
$$ \sum_{k=2}^{K}\int_{0}^{t}h^{(k)}(s)\mathrm{d}\overline{X}^{(k)}(s) \text{ } \text{ } \text{ }  \text{and }  \text{ } \text{ } \sum_{l=1}^{K}\int_{0}^{t} h^{(l)}_{i }(s)\mathrm{d}\overline{\Psi}_{i}^{(l)}(s)    $$
in $ L^{2} (\Omega, \mathcal{F}, \mathbb{Q})$. Since $L^{2} (\Omega, \mathcal{F}, \mathbb{Q})$ is a Hilbert space, the right-hand side of  (\ref{Theorem7}) is understood as the limit of the above expressions in $L^{2} (\Omega, \mathcal{F}, \mathbb{Q})$ as $K \rightarrow \infty$.
\end{remark}
\text { } \text { } We are ready to prove the main result of this section.
\begin{theorem}\label{98}
The market (\ref{55}) under $\mathbb{Q}$ is asymptotically complete.
\end{theorem}
\proof
 We consider a square-integrable contingent claim $A$ with maturity $T$. Let
$$M(t):=E_{ \mathbb{Q}}\left[\exp \left(-\int_{0}^{T}r(s)\mathrm{d}s\right) A  \bigg\arrowvert \mathcal{F}_{t}\right] $$
and
\begin{eqnarray}\label{8888}
M^{K}(t) &:=& M^{K}(0)+ \int_{0}^{t} h_{0}(s)\mathrm{d}X^{\mathbb{Q}}(s)+ \sum_{j=1}^{N}\int_{0}^{t}h_{j}(s)\mathrm{d}\overline{\Phi}_{j}^{\mathbb{Q}}(s)+ \sum_{k=2}^{K}\int_{0}^{t}h^{(k)}(s)\mathrm{d}\overline{X}^{(k)}(s)\\
\nonumber & & +\sum_{i=1}^{N}\sum_{l=1}^{K}\int_{0}^{t} h^{(l)}_{i }(s)\mathrm{d}\overline{\Psi}_{i}^{(l)}(s).
\end{eqnarray}
By the Martingale Representation Property given in Theorem \ref{94} we see that
\begin{equation}\label{K8}
\lim_{K\to \infty} M^{K}(t)=M(t)
\end{equation}
in $ L^{2} (\Omega, \mathcal{F}, \mathbb{Q})$.
For $K\geq 2$ we introduce the sequence of portfolios
\begin{equation*}
\theta^{K}(t):=\Big(\alpha^{K}(t),\beta_{0}(t),\beta_{1}(t), \ldots ,\beta_{N}(t),\beta^{(2)}(t), \ldots ,\beta^{(K)}(t),\beta_{1}^{(1)}(t), \ldots ,\beta_{N}^{(K)}(t)\Big).
\end{equation*}
We assume that all processes in $\theta^{K}$ are predictable and
$$ \int_{0}^{t} \big( \alpha^{K} (s)  \big)^2 \mathrm{d}s <\infty,  \text{  } \text{  } \int_{0}^{t} \big(\beta_{0}(s) \big)^2 \mathrm{d}\langle S_0 \rangle(s) <\infty, \text{  } \text{  }   \int_{0}^{t} \big(\beta_{j}(s) \big)^2 \mathrm{d}\langle S_j \rangle(s) <\infty, $$
$$  \int_{0}^{t} \big(\beta^{(k)}(s) \big)^2 \mathrm{d}\langle S^{(k)} \rangle(s) <\infty, \text{  } \text{  }  \int_{0}^{t} \big(\beta_{i}^{(l)}(s) \big)^2 \mathrm{d}\langle S_{i}^{(l)} \rangle(s) <\infty.  $$
Here $\alpha^{K}$ corresponds to the number of risk-free assets, $\beta_{0}$ is the number of stocks, $\beta_{j}$ $(j=1, \ldots ,N)$ is the number of units of the $j$th geometric Markovian jump security, $\beta^{(k)}$ $(k= 2, \ldots, K)$ is the number of assets $ S^{(k)}$, and $\beta_{i}^{(l)}$ $(i=1, \ldots ,N, l= 1, \ldots,K)$ is the number of assets $ S_{i}^{(l)}$.


We construct the portfolio $\theta^{K}$ as follows:
\begin{eqnarray}\label{99}
\nonumber \alpha^{K}(t)& := &M^{K}(t-)-\beta_{0}(t)B^{-1}(t)S_{0}(t-)-\sum_{j=1}^{N}\beta_{j}(t)B^{-1}(t)S_{j}(t-)\\
\nonumber &-& \sum_{k=2}^{K}\beta^{(k)}(t)B^{-1}(t)S^{(k)}(t-)- \sum_{i=1}^{N}\sum_{l=1}^{K}\beta_{i}^{(l)}(t)B^{-1}(t)S_{i}^{(l)}(t-),\\
\nonumber \beta_{0}(t)& := &h_{0}(t)B(t)S_{0}^{-1}(t-),\\
\beta_{j}(t)& := &\frac{h_{j}(t)}{\sigma_{j}(t-)}B(t)S_{j}^{-1}(t-),\\
\nonumber \beta^{(k)}(t)& := &\frac{h^{(k)}(t)}{\sigma^{(k)}(t-)}B(t)(S^{(k)})^{-1}(t-),\\
\nonumber \beta_{i}^{(l)}(t)& := &\frac{h_{i}^{(l)}(t)}{\sigma_{i}^{(l)}(t-)}B(t)(S_{i}^{(l)})^{-1}(t-).
\end{eqnarray}
The proof shows that all stochastic integrals  for this portfolio are well-defined.\\
Note that,
\begin{eqnarray}\label{delta1}
\nonumber  \Delta M^{K}(t)& = &h_{0}(t)\Delta X^{\mathbb{Q}}(t)+ \sum_{j=1}^{N} h_{j}(t)\Delta\overline{\Phi}^{\mathbb{Q}}_{j}(t)+\sum_{k=2}^{K}h^{(k)}(t)\Delta \overline{X}^{(k)}(t)+\sum_{i=1}^{N}\sum_{l=1}^{K}h^{(l)}_{i }(t)\Delta \overline{\Psi}_{i}^{(l)}(t),\\
  \Delta S_{0}(t)& = &S_{0}(t-)\Delta X^{\mathbb{Q}}(t),\\
\nonumber  \Delta S_{j}(t)& = &S_{j}(t-)\sigma_{j}(t-)\Delta \overline{\Phi}_{j}^{\mathbb{Q}} (t),\\
\nonumber  \Delta S^{(k)}(t)& = &S^{(k)}(t-)\sigma^{(k)}(t-)\Delta \overline{X}^{(k)}(t),\\
\nonumber  \Delta S_{i}^{(l)}(t)& = &S_{i}^{(l)}(t-)\sigma_{i}^{(l)}(t-)\Delta \overline{\Psi}_{i}^{(l)}(t).
\end{eqnarray}
We claim that $\{\theta^{K}, K\geq 2\}$ is the sequence of self-financing portfolios which replicates $A$. Indeed, by (\ref{99}) and (\ref{delta1}), the value $V^{K}$ of the portfolio $\theta^{K}$ is expressed by
\begin{eqnarray*}
V^{K}(t) & = & \alpha^{K}(t)B(t)+\beta_{0}(t)S_{0}(t)+\sum_{j=1}^{N} \beta_{j}(t)S_{j}(t)+\sum_{k=2}^{K}\beta^{(k)}(t)S^{(k)}(t)
+\sum_{i=1}^{N}\sum_{l=1}^{K}\beta_{i}^{(l)}(t)S_{i}^{(l)}(t)\\
\nonumber & = & M^{K}(t)B(t)-\Delta M^{K}(t)B(t)+\beta_{0}(t)\Delta  S_{0}(t)+\sum_{j=1}^{N} \beta_{j}(t)\Delta S_{j}(t) +\sum_{k=2}^{K}\beta^{(k)}(t)\Delta S^{(k)}(t)\\
\nonumber & & + \sum_{i=1}^{N}\sum_{l=1}^{K}\beta_{i}^{(l)}(t)\Delta S_{i}^{(l)}(t)  =  M^{K}(t)B(t).
\end{eqnarray*}
Thus the sequence of portfolios $\{\theta^{K}, K\geq 2\}$ replicates the claim $A$. We denote
\begin{eqnarray}\label{64}
G^{K}(u) &:=& \int_{0}^{u} \alpha^{K}(t)\mathrm{d}B(t) +\int_{0}^{u} \beta_{0}(t)\mathrm{d}S_{0}(t) +\sum_{j=1}^{N}\int_{0}^{u} \beta_{j}(t)\mathrm{d}S_{j}(t)\\
\nonumber & & {}  + \sum_{k=2}^{K}\int_{0}^{u}\beta^{(k)}(t)\mathrm{d}S^{(k)}(t)+\sum_{i=1}^{N}\sum_{l=1}^{K}\int_{0}^{u} \beta_{i}^{(l)}(t)\mathrm{d}S_{i}^{(l)}(t)
\end{eqnarray}
the gain process, i.e. the gains or losses obtained up to time $u$ by following $\theta^{K}$. We will show
\begin{equation}\label{611}
G^{K}(u)+M^{K}(0)=M^{K}(u)B(u),
\end{equation}
which implies that the portfolio is self-financing. Note that from  (\ref{K8}) we have
\begin{equation*}
\lim_{K\to \infty} G^{K}(u)=\lim_{K\to \infty}M^{K}(u)B(u)-\lim_{K\to \infty}M^{K}(0)= M(u)B(u)-M(0).
\end{equation*}
Thus the portfolios with infinitely many assets are self-financing as well. Inserting equations (\ref{99}) into (\ref{64}) we derive
\begin{eqnarray}\label{g}
\lefteqn{G^{K}(u)  =  \int_{0}^{u} M^{K}(t-)\mathrm{d}B(t)-\int_{0}^{u} h_{0}(t)\mathrm{d}B(t)-\sum_{k=2}^{K}\int_{0}^{u} \frac{h^{(k)}(t)}{\sigma^{(k)}(t-)}\mathrm{d}B(t) \nonumber}\\
&- & \sum_{j=1}^{N}\int_{0}^{u}\frac{h_{j}(t)}{\sigma_{j}(t-)}\mathrm{d}B(t)- \sum_{i=1}^{N}\sum_{l=1}^{K} \int_{0}^{u}\frac{h_{i}^{(l)}(t)}{\sigma_{i}^{(l)}(t-)}\mathrm{d}B(t)+ \int_{0}^{u}h_{0}(t)B(t)S_{0}^{-1}(t-)\mathrm{d}S_{0}(t)\\
\nonumber &+ & \sum_{j=1}^{N}\int_{0}^{u}\frac{h_{j}(t)}{\sigma_{j}(t-)}B(t)S_{j}^{-1}(t-)\mathrm{d}S_{j}(t)+ \sum_{k=2}^{K}\int_{0}^{u}\frac{h^{(k)}(t)}{\sigma^{(k)}(t-)}B(t)(S^{(k)})^{-1}(t-)\mathrm{d}S^{(k)}(t)\\
\nonumber &+ &\sum_{i=1}^{N}\sum_{l=1}^{K} \int_{0}^{u}\frac{h_{i}^{(l)}(t)}{\sigma_{i}^{(l)}(t-)}B(t)(S_{i}^{(l)})^{-1}(t-)\mathrm{d}S_{i}^{(l)}(t).
\end{eqnarray}
From the Martingale Representation Property given in Theorem \ref{94}, the first component of the above sum has the form
\begin{eqnarray*}
\lefteqn{\int_{0}^{u} M^{K}(t-)\mathrm{d}B(t)  =  \int_{0}^{u} \bigg(M^{K}(0)+ \int_{0}^{t-} h_{0}(s)\mathrm{d}X^{\mathbb{Q}}(s)+ \sum_{j=1}^{N}\int_{0}^{t-}h_{j}(s)\mathrm{d}\overline{\Phi}_{j}^{\mathbb{Q}}(s)}\\
& & +\sum_{k=2}^{K}\int_{0}^{t-}h^{(k)}(s)\mathrm{d}\overline{X}^{(k)}(s)+\sum_{i=1}^{N}\sum_{l=1}^{K}\int_{0}^{t-} h^{(l)}_{i }(s)\mathrm{d}\overline{\Psi}_{i}^{(l)}(s)\bigg)\mathrm{d}B(t)\\
& = & M^{K}(0)(B(u)-B(0))+\int_{0}^{u} h_{0}(s)(B(u)-B(s))\mathrm{d}X^{\mathbb{Q}}(s)+\sum_{j=1}^{N}\int_{0}^{u} h_{j}(s)(B(u)-B(s))\mathrm{d}\overline{\Phi}_{j}^{\mathbb{Q}}(s)\\
& & +\sum_{k=2}^{K}\int_{0}^{u} h^{(k)}(s)(B(u)-B(s))\mathrm{d}\overline{X}^{(k)}(s)+\sum_{i=1}^{N}\sum_{l=1}^{K}\int_{0}^{u} h^{(l)}_{i }(s)(B(u)-B(s))\mathrm{d}\overline{\Psi}_{i}^{(l)}(s).
\end{eqnarray*}

Now, using equation (\ref{8888}) and fact that $B(0)=1$, we can rewrite the above as follows:

\begin{eqnarray*}
\lefteqn{\int_{0}^{u} M^{K}(t-)\mathrm{d}B(t)  =  M^{K}(0)(B(u)-B(0))+B(u)(M^{K}(u)-M^{K}(0))
 -\int_{0}^{u} h_{0}(s)B(s)\mathrm{d}\overline{X}^{\mathbb{Q}}(s)}\\
& & {}- \sum_{j=1}^{N}\int_{0}^{u} h_{j}(s)B(s)\mathrm{d}\overline{\Phi}_{j}^{\mathbb{Q}}(s)
 -\sum_{k=2}^{K}\int_{0}^{u} h^{(k)}(s)B(s)\mathrm{d}\overline{X}^{(k)}(s)-\sum_{i=1}^{N}\sum_{l=1}^{K}\int_{0}^{u} h^{(l)}_{i }(s)B(s)\mathrm{d}\overline{\Psi}_{i}^{(l)}(s)\\
 & = & M^{K}(u)B(u)-M^{K}(0)-\int_{0}^{u} h_{0}(s)B(s)\mathrm{d}X^{\mathbb{Q}}(s)-\sum_{j=1}^{N}\int_{0}^{u} h_{j}(s)B(s)\mathrm{d}\overline{\Phi}_{j}^{\mathbb{Q}}(s)\\
 & & -\sum_{k=2}^{K}\int_{0}^{u} h^{(k)}(s)B(s)\mathrm{d}\overline{X}^{(k)}(s)-\sum_{i=1}^{N}\sum_{l=1}^{K}\int_{0}^{u} h^{(l)}_{i }(s)B(s)\mathrm{d}\overline{\Psi}_{i}^{(l)}(s).
\end{eqnarray*}
Inserting the above  equality into (\ref{g}), the gain process can be written as:
\begin{eqnarray*}
\lefteqn{G^{K}(u) = M^{K}(u)B(u)-M^{K}(0)-\int_{0}^{u} h_{0}(t)B(t)\mathrm{d}X^{\mathbb{Q}}(t)-\sum_{j=1}^{N}\int_{0}^{u} h_{j}(t)B(t) \mathrm{d}\overline{\Phi}_{j}^{\mathbb{Q}}(t)}\\
& & {} -\sum_{k=2}^{K}\int_{0}^{u} h^{(k)}(t)B(t)\mathrm{d}\overline{X}^{(k)}(t)-\sum_{i=1}^{N}\sum_{l=1}^{K}\int_{0}^{u} h^{(l)}_{i }(t)B(t)\mathrm{d}\overline{\Psi}_{i}^{(l)}(t)-\int_{0}^{u} h_{0}(t)\mathrm{d}B(t)\\
& & {} -\sum_{k=2}^{K}\int_{0}^{u} \frac{h^{(k)}(t)}{\sigma^{(k)}(t-)}\mathrm{d}B(t)- \sum_{j=1}^{N}\int_{0}^{u}\frac{h_{j}(t)}{\sigma_{j}(t-)}\mathrm{d}B(t) -\sum_{i=1}^{N}\sum_{l=1}^{K} \int_{0}^{u}\frac{h_{i}^{(l)}(t)}{\sigma_{i}^{(l)}(t-)}\mathrm{d}B(t)\\
& & {} +\int_{0}^{u}h_{0}(t)B(t)S_{0}^{-1}(t-)\mathrm{d}S_{0}(t)+ \sum_{j=1}^{N}\int_{0}^{u}\frac{h_{j}(t)}{\sigma_{j}(t-)}B(t)S_{j}^{-1}(t-)\mathrm{d}S_{j}(t)\\
& & {}  +\sum_{k=2}^{K}\int_{0}^{u}\frac{h^{(k)}(t)}{\sigma^{(k)}(t-)}B(t)(S^{(k)})^{-1}(t-)\mathrm{d}S^{(k)}(t)+\sum_{i=1}^{N}\sum_{l=1}^{K} \int_{0}^{u}\frac{h_{i}^{(l)}(t)}{\sigma_{i}^{(l)}(t-)}B(t)(S_{i}^{(l)})^{-1}(t-)\mathrm{d}S_{i}^{(l)}(t)\\
& =& M^{K}(u)B(u)-M(0).
\end{eqnarray*}
Thus equation (\ref{611}) holds true and the portfolio $\theta^{K}$ is self-financing.\\
\halmos
\vb

\section{Optimal portfolio selection in an It\^{o}-Markov additive market}

\text { } \text { } In this section we solve the optimization problem related to identifying the optimal strategy that maximizes the expected value of the utility function of the wealth process at the end of some fixed period. The analysis is conducted for the logarithmic and power utility functions.\\
\text { } \text { } Recall that our It\^{o}-Markov additive market is given by (\ref{47}). Equations (\ref{46}) and (\ref{477}) can be rewritten as follows:
\begin{eqnarray*}
\mathrm{d}S^{(k)}(t)&= &S^{(k)}(t-)\bigg[\mu^{(k)}(t)\mathrm{d}t+\int_{\mathbb{R}}\sigma^{(k)}(t-)\gamma^{k} (t-,x) \bar{\Pi}(\mathrm{d}t,\mathrm{d}x)\bigg],\\
\mathrm{d}S_{i}^{(l)}(t)&=&S_{i}^{(l)}(t-)\bigg[\mu_{i}^{(l)}(t)\mathrm{d}t+\int_{\mathbb{R}}x^{l} \sigma_{i}^{(l)}(t-)\bar{\Pi}_{U}^{i}(\mathrm{d}t,\mathrm{d}x)\bigg]
\end{eqnarray*}
for $i=1, \ldots ,N $, $k \geq 2$ and $l \geq 1$. Note that we consider the price processes with respect the original probability measure $\mathbb{P}$.\\
\text { } \text { }  We consider an agent who invests his initial wealth in financial assets in our market in order to maximize the expected utility of the terminal wealth.
We restrict ourselves to self-financing portfolio strategies.
Denote by ${\pi}_{0}$ the  proportion of wealth invested in stock. Let ${\pi}_{j}$ $(j=1, \ldots ,N)$, ${\pi}^{(k)}$  $(k \geq 2)$ and $\pi^{(l)}_{i}$ $(i=1,\ldots,N,  l \geq 1)$ be the proportions of wealth invested in the $j$th geometric Markovian jump security $S_j$, in the Markovian power-jump securities $S^{(k)}$ and in the impulse regime switching securities $S^{(l)}_{i}$, respectively. The balance of the investor's wealth is invested in the risk-free asset. We denote by ${\pi}(t)=({\pi}_{0}(t), {\pi}_{1}(t), \ldots , {\pi}_{N}(t), \pi^{(2)}(t),  \ldots , \pi_{1}^{(1)}(t),  \pi_{2}^{(1)}(t), \ldots)$ a portfolio strategy. We do allow short selling, but we assume that the wealth process is nonnegative at any instant (see Tepl\'a \cite{Te}).  \\
\text { } \text { } Let $K<\infty$ be the number of different assets held by the investor in his portfolio.
The wealth process $ R_{\pi}^K$ for the first $K$ assets is governed by the following stochastic differential equation (for $t\in [0,T]$):
\begin{eqnarray} \label{Rpi}
\dfrac{\mathrm{d}R_{{\pi}}^K(t)}{R_{{\pi}}^K(t-)} & := & \bigg( r(t) +\sum_{j=0}^{N}\pi_{j}(t)\big(\mu_{j}(t)-r(t)\big) +
\sum_{k=2}^{K}\pi^{(k)}(t)\big(\mu^{(k)} (t)-r(t)\big)\nonumber\\
& & {} +\sum_{i=1}^{N} \sum_{l=1}^{K}\pi^{(l)}_{i}(t)\big(\mu^{(l)}_{i}(t)-r(t)\big)
\bigg)\mathrm{d}t +\pi_{0}(t) \sigma_{0}(t-)\mathrm{d}W(t) +\sum_{j=1}^{N}\pi_{j}(t)\sigma_{j}(t-)\mathrm{d}\overline{\Phi}_{j}(t)\\
\nonumber  & & {} +\int_{\mathbb{R}}\bigg(\pi_{0}(t)\gamma (t-,x)+ \sum_{k=2}^{K}\pi^{(k)}(t)\sigma^{(k)}(t-) \gamma^{k} (t-,x)\bigg) \bar{\Pi}(\mathrm{d}t,\mathrm{d}x)\\
\nonumber & & {} +\sum_{i=1}^{N} \int_{\mathbb{R}} \bigg(x \pi_{0}(t)+ \sum_{l=1}^{K} x^{l} \pi^{(l)}_{i}(t)\sigma^{(l)}_{i}(t-) \bigg) \bar{\Pi}_{U}^{i}(\mathrm{d}t,\mathrm{d}x).
\end{eqnarray}

Note that in (\ref{Rpi}) we can take $r(t)=r(t-)$, $\mu_{j}(t)=\mu_{j}(t-)$  $(j=0,1, \ldots ,N)$, $\mu^{(k)}(t)=\mu^{(k)}(t-)$ $(k \geq 2)$ and $\mu_{i}^{(l)}(t)=\mu_{i}^{(l)}(t-)$  $(i=1,\ldots,N,  l \geq 1)$. \\
\text { } \text { }  Let $\mathcal{A}$ be the class of admissible portfolio strategies $\pi $ such that $\pi $ is predictable,\linebreak
 $ R_{\pi}^K>0,$
 $\int\limits_{0}^{T}|\pi(t)|^{2} \mathrm{d}t<\infty \text{ } \text{ } \mathbb{P}- \rm{a. s.}$,
 $\int\limits_{\mathbb{R}}\big(\pi_{0}(t)\gamma (t-,x)+ \sum_{k=2}^{K}\pi^{(k)}(t)\sigma^{(k)}(t-) \gamma^{k} (t-,x)\big) \bar{\Pi}(\mathrm{d}t,\mathrm{d}x)<\infty $,
 $\int\limits_{\mathbb{R}} \sum_{i=1}^{N}  \big(x \pi_{0}(t)+ \sum_{l=1}^{K} x^{l} \pi^{(l)}_{i}(t)\sigma^{(l)}_{i}(t-) \big) \bar{\Pi}_{U}^{i}(\mathrm{d}t,\mathrm{d}x)<\infty $
 and $\pi $ satisfies the following convergence: the wealth process $R_{\pi}^{K}$ converges to a process $R_{\pi}$ in $L^2 (\Omega, \mathcal{F}, \mathbb{P})$, where $R_{\pi}^{K}$ is the solution of the SDE (\ref{Rpi})
 (see It\^{o}'s formula in Protter \cite[Thm. 32]{P}), that is,
\begin{eqnarray*}\label{1222}
\nonumber R^{K}_{{\pi}}(t) &=& R^{K}_{{\pi}}(0)\exp \Bigg[\int_{0}^{t} \bigg( r(s-) +\sum_{j=0}^{N}\pi_{j}(s)(\mu_{j}(s-)-r(s-))+ \sum_{k=2}^{K}\pi^{(k)}(s)\big(\mu^{(k)} (s)-r(s-)\big) \\
& & + \sum_{i=1}^{N} \sum_{l=1}^{K}\pi^{(l)}_{i}(s)\big(\mu^{(l)}_{i}(s)-r(s-)\big)-\frac{1}{2}\pi_{0}^{2}(s) \sigma_{0}^{2}(s-)\bigg)\mathrm{d}s\\
\nonumber & & + \int_{0}^{t}\pi_{0}(s) \sigma_{0}(s-)dW(s)+  \sum_{j=1}^{N}\int_{0}^{t}\bigg(\log\big( 1 + \pi_{j}(s)\sigma_{j}(s-) \big) - \pi_{j}(s)\sigma_{j}(s-) \bigg) \lambda_{j}(s)ds\\
\nonumber & & + \int_{0}^{t}\int_{\mathbb{R}} \log\bigg( 1+ \pi_{0}(s)\gamma (s-,x)+ \sum_{k=2}^{K}\pi^{(k)}(s)\sigma^{(k)}(s-) \gamma^{k} (s-,x) \bigg) \bar{\Pi}(\mathrm{d}s,\mathrm{d}x)\\
\nonumber & &  + \int_{0}^{t}\int_{\mathbb{R}} \bigg(\log\Big(1+\pi_{0}(s)\gamma (s-,x)+ \sum_{k=2}^{K}\pi^{(k)}(s)\sigma^{(k)}(s-) \gamma^{k} (s-,x) \Big) -\pi_{0}(s)\gamma (s-,x)\\
\nonumber & &  - \sum_{k=2}^{K}\pi^{(k)}(s)\sigma^{(k)}(s-) \gamma^{k} (s-,x)\bigg)\nu (\mathrm{d}x)\mathrm{d}s + \sum_{j=1}^{N}\int_{0}^{t}\log\big( 1+ \pi_{j}(s)\sigma_{j}(s-) \big) \mathrm{d}\overline{\Phi}_{j}(s)\\
\nonumber & &  + \sum_{i=1}^{N} \int_{0}^{t}\int_{\mathbb{R}} \log\bigg(1+x \pi_{0}(s)+\sum_{l=1}^{K} \pi^{(l)}_{i}(s)\sigma^{(l)}_{i}(s-) x^{l} \bigg)\bar{\Pi}_{U}^{i}(\mathrm{d}s,\mathrm{d}x)\\
\nonumber & &  + \sum_{i=1}^{N} \int_{0}^{t}\int_{\mathbb{R}}\bigg( \log\Big(1+x \pi_{0}(s)+\sum_{l=1}^{K} \pi^{(l)}_{i}(s)\sigma^{(l)}_{i}(s-) x^{l}\Big)-x \pi_{0}(s)\\
\nonumber & &  - \sum_{l=1}^{K} \pi^{(l)}_{i}(s)\sigma^{(l)}_{i}(s-) x^{l}\bigg)\lambda_{i}(s) \eta (\mathrm{d}x)\mathrm{d}s \Bigg].
\end{eqnarray*}
In other words, for $\pi\in\mathcal{A}$ we require that,
\begin{equation} \label{conv4}
\lim\limits_{K \to \infty } R_{\pi}^{K}(t)=R_{\pi}(t)
\end{equation}
in $L^2 (\Omega, \mathcal{F}, \mathbb{P})$.

\begin{remark}\label{rem34}\rm
Note that (\ref{conv4}) holds true if $$\mathbb{E}_{t,z,i} \big(R_{\pi}^{K}(t)\big)^2<\infty, \text{ }  \text{ }   \mathbb{E}_{t,z,i} \big(R_{\pi}(t)\big)^2<\infty,$$
and
$$\mathbb{E}_{t,z,i}  \bigg\arrowvert   \int_{0}^{t}  \sum_{k=K+1}^{\infty}\pi^{(k)}(s)\big(\mu^{(k)} (s)-r(s-)\big) \mathrm{d}s  \bigg\arrowvert, $$
$$\mathbb{E}_{t,z,i}  \bigg\arrowvert   \int_{0}^{t} \sum_{i=1}^{N} \sum_{l=K+1}^{\infty}\pi^{(l)}_{i}(s)\big(\mu^{(l)}_{i}(s)-r(s-)\big)  \mathrm{d}s      \bigg\arrowvert ,$$
$$\mathbb{E}_{t,z,i}  \bigg\arrowvert  \int_{0}^{t}\int_{\mathbb{R}} \log\bigg( 1+ \pi_{0}(s)\gamma (s-,x)+ \sum_{k=K+1}^{\infty}\pi^{(k)}(s)\sigma^{(k)}(s-) \gamma^{k} (s-,x) \bigg) \bar{\Pi}(\mathrm{d}s,\mathrm{d}x)        \bigg\arrowvert^2 ,$$
$$\mathbb{E}_{t,z,i}  \bigg\arrowvert   \int_{0}^{t}\int_{\mathbb{R}} \bigg(\log\Big(1+\pi_{0}(s)\gamma (s-,x)+ \sum_{k=K+1}^{\infty}\pi^{(k)}(s)\sigma^{(k)}(s-) \gamma^{k} (s-,x) \Big) -\pi_{0}(s)\gamma (s-,x)$$
$$  - \sum_{k=K+1}^{\infty}\pi^{(k)}(s)\sigma^{(k)}(s-) \gamma^{k} (s-,x)\bigg)\nu (\mathrm{d}x)\mathrm{d}s   \bigg\arrowvert ,$$
$$\mathbb{E}_{t,z,i}  \bigg\arrowvert   \int_{0}^{t}\int_{\mathbb{R}} \log\bigg(1+x \pi_{0}(s)+\sum_{l=K+1}^{\infty} \pi^{(l)}_{i}(s)\sigma^{(l)}_{i}(s-) x^{l} \bigg)\bar{\Pi}_{U}^{i}(\mathrm{d}s,\mathrm{d}x) \bigg\arrowvert^2, $$
$$\mathbb{E}_{t,z,i}  \bigg\arrowvert   \sum_{i=1}^{N} \int_{0}^{t}\int_{\mathbb{R}}\bigg( \log\Big(1+x \pi_{0}(s)+\sum_{l=K+1}^{\infty} \pi^{(l)}_{i}(s)\sigma^{(l)}_{i}(s-) x^{l}\Big)-x \pi_{0}(s) $$
$$- \sum_{l=K+1}^{\infty} \pi^{(l)}_{i}(s)\sigma^{(l)}_{i}(s-) x^{l}\bigg)\lambda_{i}(s) \eta (\mathrm{d}x)\mathrm{d}s     \bigg\arrowvert $$
tend to $0$ as $K \rightarrow \infty$.
Indeed, the convergence (\ref{conv4}) follows directly from our assumptions:
from the triangle inequality and the inequality (see Fechner \cite{F})
$$2\arrowvert \exp(a_1) -\exp(a_2) \arrowvert \leq \arrowvert a_1-a_2 \arrowvert \arrowvert \exp(a_1) + \exp(a_2) \arrowvert  \text {, }  \text { }  \text { } a_1, a_2 \in \mathbb{R}, $$
we get
$$\mathbb{E} \big\arrowvert R_{\pi}(t)- R_{\pi}^{K}(t) \big\arrowvert^2 \leq   \mathbb{E}  \bigg\arrowvert  \log  \dfrac{R_{{\pi}}(t)}{R_{{\pi}}^K(t)}  \bigg\arrowvert^2  \Big(\mathbb{E} \arrowvert R_{\pi}(t)\arrowvert^2  + \mathbb{E} \arrowvert R_{\pi}^{K}(t)\arrowvert^2  \Big)<\infty. $$
\end{remark}
\text { } \text { }  Let $U $ denote a utility function of the investor, which is strictly increasing, strictly concave and twice differentiable, that is, $U'>0 $ and $U''<0 $.

For each $(t,z)\in \mathbb{T}\times \mathbb{R}^{+} $ and each $i=1, \ldots ,N$ we define
$$V^{\pi}(t,z,\textbf{e}_{i}):=\mathbb{E}_{t,z,i}\big[U(R_{\pi}(T))\big], $$
where $\mathbb{E}_{t,z,i}$ is the conditional expectation given $R_{\pi}(0)=z$ and $J(t) = \textbf{e}_{i}$ under $\mathbb{P}$. \\
The expectation above is understood in the limiting sense, that is, we limit the set of admissible strategies $\mathcal{A}$ to the strategies $\pi$ such that $ \lim\limits_{K \to \infty  } \mathbb{E}_{t,z,i}\big[U\big(R_{\pi}^{K}(T)\big)\big]$ exists and is finite. In other words,
\begin{equation}\label{lim2}
 \lim\limits_{K \to \infty  } \mathbb{E}_{t,z,i}\big[U\big(R_{\pi}^{K}(T)\big)\big]= \mathbb{E}_{t,z,i}\big[U \big(R_{\pi}(T)\big)\big]<\infty.
\end{equation}

Then the value function of the investor's portfolio selection problem is defined by
\begin{equation} \label{1136}
V(t,z,\textbf{e}_{i}):=\sup\limits_{\pi\in\mathcal{A} }V^{\pi}(t,z,\textbf{e}_{i})= \sup\limits_{\pi\in\mathcal{A} }\mathbb{E}_{t,z,i}\big[U(R_{\pi}(T))\big].
\end{equation}

\begin{lemma} \label{lemma1}
Under assumption (\ref{conv4}) equation (\ref{lim2}) holds true.
\end{lemma}

\proof
We define
\begin{equation*}\label{UM}
U_M (z):=U(z) \mathbbm{1}_{\{z: |U(z)| \leq M \}}.
\end{equation*}
The convergence of $R_{\pi}^{K}$ to $R_{\pi}$ in $L^2 (\Omega, \mathcal{F}, \mathbb{P})$ given in (\ref{conv4}) implies
the convergence in probability of $R_{\pi}^{K}$ to $R_{\pi}$ as $K \rightarrow \infty$ (see Jacod and Protter \cite[Thm. 17.2]{JaPr}). \\
Thus, for the bounded and continuous function $U_M$ given in (\ref{UM}) we have
\begin{equation*}\label{W1}
\lim\limits_{K \to \infty  } \mathbb{E}_{t,z,i}\big[U_M \big(R_{\pi}^{K}(T)\big)\big]= \mathbb{E}_{t,z,i}\big[U_M \big(R_{\pi}(T)\big)\big]
\end{equation*}
(see Jacod and Protter \cite[Thm. 18.1]{JaPr}). \\
Note that
\begin{equation}\label{URKpi}
 \mathbb{E}_{t,z,i}\big[U \big(R_{\pi}^{K}(T)\big)\big]= \mathbb{E}_{t,z,i}\big[U_M \big(R_{\pi}^{K}(T)\big)\big]+ \mathbb{E}_{t,z,i}\big[U \big(R_{\pi}^{K}(T)\big)\mathbbm{1}_{\{R_{\pi}^{K}(T):  |U(R_{\pi}^{K}(T))| > M \}}\big] .
\end{equation}
From the concavity of $U$ it follows that $U(z)\leq b+c z$ for each $z\geq 0$ and some real  $b,c\geq 0$.
Thus the second term on the right-hand side in (\ref{URKpi}) satisfies
\begin{eqnarray*}\label{}
\mathbb{E}_{t,z,i}\big[U \big(R_{\pi}^{K}(T)\big) \mathbbm{1}_{\{R_{\pi}^{K}(T):  |U(R_{\pi}^{K}(T))| > M \}}\big] &\leq &
b+c \text{ } \mathbb{E}_{t,z,i}\big[R_{\pi}^{K}(T)\mathbbm{1}_{\{R_{\pi}^{K}(T):  |U(R_{\pi}^{K}(T))| > M \}}\big].
 \end{eqnarray*}
Now, we will prove that
\begin{equation}\label{W2}
\lim\limits_{K \to \infty  } \mathbb{E}_{t,z,i}\big[ R_{\pi}^{K}(T)\mathbbm{ I}^K(T)\big]= \mathbb{E}_{t,z,i}\big[ R_{\pi}(T)\mathbbm{ I}(T)\big],
\end{equation}
where
\begin{equation*}\label{}
\mathbbm{ I}^K(T):=\mathbbm{1}_{\{R_{\pi}^{K}(T):  |U(R_{\pi}^{K}(T))| > M \}}
\text{ } \text{ and } \text{ }
\mathbbm{ I}(T):=\mathbbm{1}_{\{R_{\pi}(T):  |U(R_{\pi}(T))| > M \}}.
\end{equation*}
Note that
\begin{eqnarray}\label{W9}
\nonumber \lefteqn{\mathbb{E}_{t,z,i}\big|R_{\pi}^{K}(T)\mathbbm{ I}^K(T)-R_{\pi}(T)\mathbbm{ I}(T)\big|} \\
  & \leq& \bigg[\mathbb{E}_{t,z,i}\big(R_{\pi}^{K}(T)- R_{\pi}(T)\big)^2\bigg]^{\frac{1}{2}} \bigg[\mathbb{E}_{t,z,i}\big(\mathbbm{ I}(T)\big)\bigg]^{\frac{1}{2}}\\
  \nonumber & &  + \text{ } \bigg[\mathbb{E}_{t,z,i}\big(R_{\pi}(T)\big)^2\bigg]^{\frac{1}{2}} \bigg[\mathbb{E}_{t,z,i}\big( \mathbbm{ I}(T)- \mathbbm{ I}^K(T)\big)^2\bigg]^{\frac{1}{2}}\\
  \nonumber & & +\text{ } \bigg[\mathbb{E}_{t,z,i}\big(R_{\pi}^K (T)\big)^2\bigg]^{\frac{1}{2}} \bigg[\mathbb{E}_{t,z,i}\big( \mathbbm{ I}^K(T )- \mathbbm{ I}(T)\big)^2\bigg]^{\frac{1}{2}} \\
 \nonumber & & + \text{ } \bigg[\mathbb{E}_{t,z,i}\big(R_{\pi} (T)\big)^2\bigg]^{\frac{1}{2}} \bigg[\mathbb{E}_{t,z,i}\big( \mathbbm{ I}^K(T)- \mathbbm{ I}(T)\big)^2\bigg]^{\frac{1}{2}}.
\end{eqnarray}

Indeed, from the triangle inequality we obtain
\begin{eqnarray}\label{W6}
\lefteqn{\mathbb{E}_{t,z,i}\big|R_{\pi}^{K}(T)\mathbbm{ I}^K(T)-R_{\pi}(T)\mathbbm{ I}(T)\big|} \\
\nonumber & \leq&  \mathbb{E}_{t,z,i}\big|R_{\pi}^{K}(T)\mathbbm{ I}^K(T)-R_{\pi}^{K}(T) \mathbbm{ I}(T)\big|+\mathbb{E}_{t,z,i}\Big|R_{\pi}^{K}(T) \mathbbm{ I}(T)-R_{\pi}(T)\mathbbm{ I}^K(T)\Big|\\
\nonumber & & +\text { }\mathbb{E}_{t,z,i}\Big|R_{\pi}(T)\mathbbm{ I}^K(T)-R_{\pi}(T)\mathbbm{ I}(T)\Big|.
\end{eqnarray}
Moreover, using H\"older's inequality we get
\begin{eqnarray}\label{W8}
\mathbb{E}_{t,z,i}\Big|R_{\pi}^{K}(T)\mathbbm{ I}^K(T)- R_{\pi}^K(T) \mathbbm{ I}(T)\Big| \leq  \bigg[\mathbb{E}_{t,z,i}\big(R_{\pi}^K (T)\big)^2\bigg]^{\frac{1}{2}} \bigg[\mathbb{E}_{t,z,i}\big( \mathbbm{ I}^K(T)- \mathbbm{ I}(T)\big)^2\bigg]^{\frac{1}{2}}
\end{eqnarray}
and
\begin{eqnarray}\label{W10}
\mathbb{E}_{t,z,i}\Big|R_{\pi}(T)\mathbbm{ I}^K(T)- R_{\pi}(T) \mathbbm{ I}(T)\Big| \leq  \bigg[\mathbb{E}_{t,z,i}\big(R_{\pi} (T)\big)^2\bigg]^{\frac{1}{2}} \bigg[\mathbb{E}_{t,z,i}\big( \mathbbm{ I}^K(T )- \mathbbm{ I}(T)\big)^2\bigg]^{\frac{1}{2}}.
\end{eqnarray}
Finally, from the triangle inequality and H\"older's inequality we derive
\begin{eqnarray}\label{W7}
\lefteqn{\mathbb{E}_{t,z,i}\Big|R_{\pi}^{K}(T)\mathbbm{ I}(T)- R_{\pi}(T) \mathbbm{ I}^K(T)\Big|} \\
\nonumber & \leq& \bigg[\mathbb{E}_{t,z,i}\big(R_{\pi}^{K}(T)- R_{\pi}(T)\big)^2\bigg]^{\frac{1}{2}} \bigg[\mathbb{E}_{t,z,i}\big(\mathbbm{ I}(T)\big)\bigg]^{\frac{1}{2}}\\
\nonumber & & + \text{ } \bigg[\mathbb{E}_{t,z,i}\big(R_{\pi}(T)\big)^2\bigg]^{\frac{1}{2}} \bigg[\mathbb{E}_{t,z,i}\big( \mathbbm{ I}(T)- \mathbbm{ I}^K(T)\big)^2\bigg]^{\frac{1}{2}}.
\end{eqnarray}
Combining (\ref{W6}), (\ref{W8}), (\ref{W10}) and  (\ref{W7})   we get the inequality (\ref{W9}).

Now, we will prove that
\begin{equation} \label{conv5}
\mathbbm{ I}^K \rightarrow \mathbbm{ I}
\end{equation}
in $L^2 (\Omega, \mathcal{F}, \mathbb{P})$ as $K \rightarrow \infty$.

First, we verify this convergence in probability.
Indeed, for any $\varepsilon>0$ there exists $\delta>0$ such that
\begin{eqnarray*}
\mathbb{P}(|\mathbbm{ I}^K(T)- \mathbbm{ I}(T)|>\varepsilon)&=&\mathbb{P}(\{U(R^K_{\pi}(T))>M, U(R_{\pi}(T))<M\}  \cup \{U(R^K_{\pi}(T))<M, U(R_{\pi}(T))>M\} )\\
& & \leq \mathbb{P}(|U(R^K_{\pi}(T))- U(R_{\pi}(T))|>\delta).
\end{eqnarray*}
The right-hand side of the above inequality tends to zero as  $K \rightarrow \infty$ since the convergence in probability of $R_{\pi}^{K}$ to $R_{\pi}$ yields the convergence in probability of $U(R_{\pi}^{K})$ to $U(R_{\pi})$ (see Jacod and Protter \cite[Thm. 17.5]{JaPr}).

Moreover, we have
$$ \mathbbm{ I}^K(T) \leq 1,$$
thus $\{\mathbbm{ I}^K\}_{K=2}^{\infty}$ is uniformly integrable and
\begin{equation*}\label{Indyk}
\lim\limits_{K \to \infty} \mathbbm{ I}^K(T)=\mathbbm{ I}(T)
\end{equation*}
in $L^2 (\Omega, \mathcal{F}, \mathbb{P})$ (see Gut \cite[Thm. 4.5, p. 216  and Thm. 5.4, p. 221 ]{Gut}) . This completes the proof of (\ref{conv5}).  \\
By (\ref{conv4}) and (\ref{conv5}), the right-hand side of inequality (\ref{W9}) tends to $0$ as $K \rightarrow \infty$.
This completes the proof of (\ref{W2}).

From (\ref{W2}) it follows that $$\overline{\lim\limits_{K \to \infty  }} \mathbb{E}_{t,z,i}\big[U \big(R_{\pi}^{K}(T)\big) \mathbbm{1}_{\{R_{\pi}^{K}(T):  |U(R_{\pi}^{K}(T))| > M \}}\big]$$ is well-defined. Thus as $M \rightarrow \infty$, the second term on the right-hand side of (\ref{URKpi}) tends to zero. Moreover, the first term converges to $\mathbb{E}_{t,z,i}\big[U \big(R_{\pi}(T)\big)\big]$.
This completes the proof.
\halmos\\

\text { } \text { } Our main goal is to identify the value function given in (\ref{1136}).
In what follows, we consider two risk-averse utility functions, namely, the logarithmic utility and the power utility.

\subsection{Logarithmic utility}
In this subsection, we derive the optimal portfolio strategy in the case of a logarithmic utility function of wealth, namely
\begin{equation*} \label{1236}
U(z)=\log(z).
\end{equation*}

Recall that in $\mathcal{A}$ we consider only the strategies for which
\begin{equation} \label{fini}
\lim\limits_{K \to \infty  } \mathbb{E}_{t,z,i}\big[ \log R^K_{\pi}(T)\big]=\mathbb{E}_{t,z,i}\big[\log R_{\pi}(T) \big]<\infty.
\end{equation}

\begin{theorem}\label{980}
Assume that there exists a solution
$${\pi}^{\star}(t):=({\pi}_{0}^{\star}(t), {\pi}_{1}^{\star}(t), \ldots, {\pi}_{N}^{\star}(t),\pi^{(2)\star}(t), \pi^{(3)\star}(t), \ldots , \pi_{1}^{(1)\star}(t),  \pi_{2}^{(1)\star}(t),  \ldots) $$
 of the following system of equations (for $i,j=1, \ldots ,N, $ $k \geq 2$ and $l \geq 1$ ):
\begin{eqnarray}\label{509}
\nonumber  r(t-)-\mu_{0}(t-)&= &  \pi^{\star}_{0}(t)\sigma_{0}^{2}(t-) +\sum_{i=1}^{N}\dfrac{r(t-)- \mu_{i}^{(1)}(t-)}{\sigma_{i}^{(1)}(t-)}+\int_{\mathbb{R}} \gamma(t-,x) \bigg(\Big(1+ \pi^{\star}_{0}(t) \gamma (t-,x)\\
 \nonumber & & +\sum_{k=2}^{\infty}\pi^{(k)\star}(t) \sigma^{(k)}(t-) \gamma^{k}(t-,x)\Big)^{-1}  -1\bigg)\nu (\mathrm{d}x), \\
  \pi_{j}^{\star}(t)&=&\dfrac{\mu_{j}(t-)-r(t-)}{\big(r(t-)-\mu_{j}(t-)\big)\sigma_{j}(t-)+ \lambda_{j}(t)\sigma_{j}^{2}(t-)},
    \end{eqnarray}
 \begin{eqnarray*}
 \nonumber    \dfrac{r(t-)-\mu^{(k)}(t-)}{\sigma^{(k)}(t-)}&=&\int_{\mathbb{R}} \gamma^{k} (t-,x) \bigg(\Big(1+ \pi^{\star}_{0}(t) \gamma (t-,x)\\
 \nonumber & & +\sum_{k=2}^{\infty}\pi^{(k)\star}(t) \sigma^{(k)}(t-) \gamma^{k}(t-,x)\Big)^{-1}  -1\bigg)\nu (\mathrm{d}x), \\
 \nonumber   \dfrac{r(t-)-\mu^{(l)}_{i}(t-)}{\sigma^{(l)}_i(t-)} &=& \int_{\mathbb{R}}x^l \bigg(\Big(1+ x \pi^{\star}_{0}(t)  +\sum_{l=1}^{\infty}\pi^{(l) \star}_i(t) \sigma^{(l)}_i(t-) x^l \Big)^{-1}  -1\bigg)\lambda_{i}(t) \eta (\mathrm{d}x),
 \end{eqnarray*}
which belongs to $\mathcal{A}$, that is, in particular, satisfies (\ref{conv4}) and (\ref{fini}). Then the optimal portfolio strategy for the portfolio selection problem (\ref{1136}) with logarithmic utility function of wealth is one of those solutions.
\end{theorem}

\proof
The conditional expectation of the logarithm of the wealth process has the following form (for $t\in[0,T)$):
\begin{eqnarray*}
\lefteqn{ \mathbb{E}_{t,z,i}\big[\log R_{\pi}(T)\big] =  \log R_{\pi}(t)+ \mathbb{E}_{t,z,i}\int_{t}^{T}\bigg[ r(s-) +\sum_{j=0}^{N}\pi_{j}(s)\big(\mu_{j}(s-)-r(s-)\big) }\\
\nonumber & & + \sum_{k=2}^{\infty}\pi^{(k)}(s)\big(\mu^{(k)} (s)-r(s-)\big)+ \sum_{i=0}^{N} \sum_{l=1}^{\infty}\pi^{(l)}_{i}(s)\big(\mu^{(l)}_{i}(s)-r(s-)\big)-\frac{1}{2}\pi_{0}^{2}(s) \sigma_{0}^{2}(s-)\\
\nonumber & & {} + \sum_{j=1}^{N}\Big(\log\big( 1 + \pi_{j}(s)\sigma_{j}(s-) \big) - \pi_{j}(s)\sigma_{j}(s-) \Big) \lambda_{j}(s) \\
\nonumber & & {} + \int_{\mathbb{R}} \bigg(\log\Big(1+\pi_{0}(s)\gamma (s-,x)+ \sum_{k=2}^{\infty}\pi^{(k)}(s)\sigma^{(k)}(s-) \gamma^{k} (s-,x) \Big)\\
\nonumber & & {} -\pi_{0}(s)\gamma (s-,x)- \sum_{k=2}^{\infty}\pi^{(k)}(s)\sigma^{(k)}(s-) \gamma^{k} (s-,x)\bigg)\nu (\mathrm{d}x) \\
\nonumber & & {} + \sum_{i=1}^{N} \int_{\mathbb{R}}\bigg( \log\Big(1+x\pi_{0}(s)+\sum_{l=1}^{\infty} \pi^{(l)}_{i}(s)\sigma^{(l)}_{i}(s-) x^{l}\Big)-x\pi_{0}(s)-\sum_{l=1}^{\infty} \pi^{(l)}_{i}(s)\sigma^{(l)}_{i}(s-) x^{l}\bigg)\lambda_{i}(s) \eta (\mathrm{d}x) \bigg]\mathrm{d}s.
\end{eqnarray*}
Therefore the optimal value function $V$ can be written as
$$V(t,z,\textbf{e}_{i})= \log(z) +\sup\limits_{\pi\in\mathcal{A} } h_{\pi}(t,\textbf{e}_{i}),$$
where
\begin{equation*}
h_{\pi}(t,\textbf{e}_{i}):=\mathbb{E}_{t,z,i}\int_{t}^{T} F\big( \pi_{0}(s), \pi_{1}(s),  \ldots, \pi_{N}(s), \pi^{(2)}(s),  \ldots ,\pi^{(1)}_{1}(s),  \ldots \big)\mathrm{d}s
\end{equation*}
for
\begin{eqnarray*}
\lefteqn{F\big( \pi_{0}(s), \pi_{1}(s), \ldots, \pi_{N}(s), \pi^{(2)}(s),  \ldots ,\pi^{(1)}_{1}(s),  \ldots \big) := r(s-) +\sum_{j=0}^{N}\pi_{j}(s)(\mu_{j}(s-)-r(s-))}\\
\nonumber & & {} + \sum_{k=2}^{\infty}\pi^{(k)}(s)\big(\mu^{(k)} (s)-r(s-)\big) + \sum_{i=0}^{N} \sum_{l=1}^{\infty}\pi^{(l)}_{i}(s)\big(\mu^{(l)}_{i}(s)-r(s-)\big)-\frac{1}{2}\pi_{0}^{2}(s) \sigma_{0}^{2}(s-)\\
\nonumber & & {}  + \sum_{j=1}^{N}\Big(\log\big( 1 + \pi_{j}(s)\sigma_{j}(s-) \big) - \pi_{j}(s)\sigma_{j}(s-) \Big) \lambda_{j}(s)\\
\nonumber & & {} + \int_{\mathbb{R}} \bigg(\log\Big(1+\pi_{0}(s)\gamma (s-,x)+ \sum_{k=2}^{\infty}\pi^{(k)}(s)\sigma^{(k)}(s-) \gamma^{k} (s-,x) \Big)\\
\end{eqnarray*}
\begin{eqnarray*}
\nonumber & & { }  -\pi_{0}(s)\gamma (s-,x)- \sum_{k=2}^{\infty}\pi^{(k)}(s)\sigma^{(k)}(s-) \gamma^{k} (s-,x)\bigg)\nu (\mathrm{d}x)\\
\nonumber & & { } + \sum_{i=1}^{N} \int_{\mathbb{R}}\bigg( \log\Big(1+x\pi_{0}(s)+\sum_{l=1}^{\infty} \pi^{(l)}_{i}(s)\sigma^{(l)}_{i}(s-) x^{l}\Big)-x\pi_{0}(s)-\sum_{l=1}^{\infty} \pi^{(l)}_{i}(s)\sigma^{(l)}_{i}(s-) x^{l}\bigg)\lambda_{i}(s) \eta (\mathrm{d}x).
\end{eqnarray*}
Thus, to determine the optimal portfolio strategy, it is sufficient to maximize $F$.
Indeed, the maximization of the function  $F\big( \pi_{0}(s), \pi_{1}(s), \ldots, \pi_{N}(s), \pi^{(2)}(s),\ldots,  \pi^{(1)}_{1}(s), \ldots \big)$ at each time point $s \in [0, T]$ maximizes the integral of $F$ on $[0, T]$.
By direct differentiation with respect to $\pi_{0}, \pi_{j}, \pi^{(k)}, \pi^{(l)}_{i} $ we obtain conditions (\ref{509}) which the optimal strategies have to satisfy. Observe that from (\ref{42}) the integrals
\begin{eqnarray*}
\nonumber & & {} \int_{\mathbb{R}} \bigg(\log\Big(1+\pi_{0}(s)\gamma (s-,x)+ \sum_{k=2}^{\infty}\pi^{(k)}(s)\sigma^{(k)}(s-) \gamma^{k} (s-,x) \Big) -\pi_{0}(s)\gamma (s-,x)\\
\nonumber & & { } - \sum_{k=2}^{\infty}\pi^{(k)}(s)\sigma^{(k)}(s-) \gamma^{k} (s-,x)\bigg)\nu (\mathrm{d}x)
\end{eqnarray*}
 and (for $i=1,\ldots,N$)
 $$ \int_{\mathbb{R}}\bigg( \log\Big(1+x\pi_{0}(s)+\sum_{l=1}^{\infty} \pi^{(l)}_{i}(s)\sigma^{(l)}_{i}(s-) x^{l}\Big)-x\pi_{0}(s)-\sum_{l=1}^{\infty} \pi^{(l)}_{i}(s)\sigma^{(l)}_{i}(s-) x^{l}\bigg)\lambda_{i}(s) \eta (\mathrm{d}x)$$
 are well-defined. Hence by the Leibniz integral rule we can interchange the above mentioned derivatives and the integrals.

\halmos

\begin{remark}
 We have not been able to prove that a solution of system (\ref{509}) exists and is unique. On a complete It\^{o}-Markov additive market we have an infinite number of assets, so the optimal portfolio strategy ${\pi}^{\star }$ is an infinite dimensional vector. The value function  (\ref{1136}) is understood in the limiting sense and therefore numerically it can be approximated by the finite strategy counterpart.
In the case of finite dimensional approximations by Kramkov and Schachermayer \cite[Thm. 2.2]{Sch} the optimal strategy exists and is unique.
\end{remark}

\subsection{Power utility}
In this subsection, we derive the optimal portfolio strategy in the case of the power utility function, namely
\begin{equation*}\label{508}
U(z)=z^{\alpha} \qquad \qquad  \text { for } \alpha \in (0,1) .
\end{equation*}
We assume that for each $\pi\in\mathcal{A}$,
\begin{eqnarray} \label{2}
\lim\limits_{K \to \infty  } \mathbb{E}_{t,z,i} \big( R^K_{\pi}(T)\big)^{\alpha}=\mathbb{E}_{t,z,i}\big(R_{\pi}(T)\big)^{\alpha}<\infty.
\end{eqnarray}

\begin{theorem}\label{980}
Assume that there exists a solution
$${\pi}^{\star}(t):=({\pi}_{0}^{\star}(t), {\pi}_{1}^{\star}(t), \ldots, {\pi}_{N}^{\star}(t),\pi^{(2)\star}(t), \pi^{(3)\star}(t), \ldots , \pi_{1}^{(1)\star}(t),  \pi_{2}^{(1)\star}(t),  \ldots) $$
 of the following system of equations $($for $i,j=1, \ldots ,N, $ $k \geq 2$ and $l \geq 1)$:
 \begin{eqnarray*}
 \nonumber r(t-) - \mu_{0} (t)&=&(\alpha-1)  \pi^{\star}_{0}(t)\sigma_{0}^{2}(t-)+\sum_{i=1}^{N}\dfrac{\mu_{i}^{(1)}(t-)-r(t-)}{\sigma_{i}^{(1)}(t-)} +\int_{\mathbb{R}}  \gamma (t-,x) \bigg(\Big(1+ \pi_{0}^{\star}(t) \gamma (t-,x) \\
\nonumber & &  +\sum_{k=2}^{\infty}\pi^{(k)\star}(t) \sigma^{(k)}(t-) \gamma^{k}(t-,x)\Big)^{\alpha-1}  -1\bigg)\nu (\mathrm{d}x),
 \end{eqnarray*}
\begin{eqnarray}\label{5091}
  \pi_{j}^{\star}(t)&=&\dfrac{\bigg(1-\frac{\mu_{j}(t-)-r(t-)}{\lambda_{i}(t)\sigma_{j}(t-)} \bigg)^{\frac{1}{\alpha-1}} -1}{\sigma_{j}(t-)},\\
\nonumber r(t-) - \mu^{(k)} (t)&=&\int_{\mathbb{R}} \sigma^{(k)}(t-)  \gamma^{k} (t-,x) \bigg(\Big(1+ \pi_{0}^{\star}(t) \gamma (t-,x)\\
\nonumber & &  +\sum_{k=2}^{\infty}\pi^{(k)\star}(t) \sigma^{(k)}(t-) \gamma^{k}(t-,x)\Big)^{\alpha-1}  -1\bigg)\nu (\mathrm{d}x), \\
\nonumber r(t-) - \mu^{(l)}_{i} (t)&=&\int_{\mathbb{R}} \sigma^{(l)}_{i}(t-) x^{l} \bigg(\Big(1+ x \pi_{0}^{\star}(t)  +\sum_{l=1}^{\infty}\pi^{\star(l)}_{i}(t) x^l \sigma^{(l)}_{i}(t-) \Big)^{\alpha-1}  -1\bigg)  \lambda_{i}(t) \eta (\mathrm{d}x)\mathrm{d}t,
\end{eqnarray}
which belongs to $\mathcal{A}$, that is, in particular, it satisfies (\ref{conv4}) and (\ref{2}). Then the optimal portfolio strategy
for the portfolio selection problem (\ref{1136}) with power utility function of wealth is one of those solutions.
\end{theorem}

\proof
From It\^{o}'s formula (see Protter \cite[Thm. 32]{P}) for the power utility function of wealth, we obtain (for $s\in[t,T]$ and $t\in[0,T]$)
\begin{eqnarray*}
\lefteqn{( R_{{\pi}}(T))^{\alpha}-( R_{{\pi}}(t))^{\alpha} = \int_t^T \alpha ( R_{{\pi}}(s))^{\alpha} \bigg( r(s-) +\sum_{j=0}^{N}\pi_{j}(s)\big(\mu_{j}(s-)-r(s-)\big) + \sum_{k=2}^{\infty}\pi^{(k)}(s)\big(\mu^{(k)} (s)-r(s-)\big) }\\
\nonumber & & + \sum_{i=0}^{N} \sum_{l=1}^{\infty}\pi^{(l)}_{i}(s)\big(\mu^{(l)}_{i}(s)-r(s-)\big)\bigg)\mathrm{d}s+ \int_t^T\alpha( R_{{\pi}}(s))^{\alpha} \pi_{0}(s) \sigma_{0}(s-)\mathrm{d}W(s) \\
\nonumber & & {} + \int_t^T \frac{1}{2}\alpha(\alpha-1)( R_{{\pi}}(s))^{\alpha}\pi_{0}^{2}(s) \sigma_{0}^{2}(s-)\mathrm{d}s + \sum_{j=1}^{N} \int_t^T\bigg(\big( R_{{\pi}}(s)+R_{{\pi}}(s)\pi_{j}(s)\sigma_{j}(s-) \big)^{\alpha} - (R_{{\pi}}(s))^{\alpha} \bigg) \mathrm{d}\overline{\Phi}_{j}(s)\\
\nonumber & & {} + \sum_{j=1}^{N} \int_t^T\bigg(\big( R_{{\pi}}(s)+R_{{\pi}}(s)\pi_{j}(s)\sigma_{j}(s-) \big)^{\alpha} - (R_{{\pi}}(s))^{\alpha} - \alpha ( R_{{\pi}}(s))^{\alpha}\pi_{j}(s)\sigma_{j}(s-) \bigg) \lambda_{j}(s)\mathrm{d}s \\
\nonumber & & {} +  \int_t^T\int_{\mathbb{R}} \Bigg(\bigg( R_{{\pi}}(s-)+R_{{\pi}}(s)\Big(\pi_{0}(s)\gamma (s-,x)+ \sum_{k=2}^{\infty}\pi^{(k)}(s)\sigma^{(k)}(s-) \gamma^{k} (s-,x) \Big)\bigg)^{\alpha}
\end{eqnarray*}
\begin{eqnarray*}
\nonumber & & {} - (R_{{\pi}}(s-))^{\alpha} - \alpha (R_{{\pi}}(s-))^{\alpha} \Big(\pi_{0}(s)\gamma (s-,x)+ \sum_{k=2}^{\infty}\pi^{(k)}(s)\sigma^{(k)}(s-) \gamma^{k} (s-,x) \Big)\Bigg) \nu (\mathrm{d}x)\mathrm{d}s\\
\nonumber & & {} +  \int_t^T\int_{\mathbb{R}} \Bigg(\bigg( R_{{\pi}}(s-)+R_{{\pi}}(s)\Big(\pi_{0}(s)\gamma (s-,x)+ \sum_{k=2}^{\infty}\pi^{(k)}(s)\sigma^{(k)}(s-) \gamma^{k} (s-,x) \Big)\bigg)^{\alpha}  - (R_{{\pi}}(s-))^{\alpha} \Bigg) \bar{\Pi}(\mathrm{d}s,\mathrm{d}x)\\
\nonumber & & {} +\sum_{i=1}^{N}  \int_t^T\int_{\mathbb{R}} \Bigg( \bigg(R_{{\pi}}(s-)+R_{{\pi}}(s) \Big(x \pi_{0}(s)+ \sum_{l=1}^{\infty} x^{l} \pi^{(l)}_{i}(s)\sigma^{(l)}_{i}(s-)\Big) \bigg)^{\alpha} - (R_{{\pi}}(s-))^{\alpha} \\
\nonumber & & {}  - \alpha (R_{{\pi}}(s-))^{\alpha} \bigg(x \pi_{0}(s)+ \sum_{l=1}^{\infty} x^{l} \pi^{(l)}_{i}(s)\sigma^{(l)}_{i}(s-)\bigg)  \Bigg) \lambda_{i}(s) \eta (\mathrm{d}x)\mathrm{d}s \\
\nonumber & & {} +\sum_{i=1}^{N}  \int_t^T\int_{\mathbb{R}} \Bigg( \bigg(R_{{\pi}}(s-)+R_{{\pi}}(s) \Big(x \pi_{0}(s)+ \sum_{l=1}^{\infty} x^{l} \pi^{(l)}_{i}(s)\sigma^{(l)}_{i}(s-)\Big) \bigg)^{\alpha} - \big(R_{{\pi}}(s-)\big)^{\alpha}  \Bigg) \bar{\Pi}_{U}^{i}(\mathrm{d}s,\mathrm{d}x).
\end{eqnarray*}
From this and (\ref{1136}) the value function is given by
\begin{eqnarray*}
V(t,z,\textbf{e}_{i})&=& z^{\alpha} + \sup\limits_{\pi\in\mathcal{A} } \mathbb{E}_{t,z,i} \int_{t}^{T} z^{\alpha}  \Bigg[ \alpha \bigg( r(s-)  +\sum_{j=0}^{N}\pi_{j}(s)\big(\mu_{j}(s-)-r(s-)\big)+
 \sum_{k=2}^{\infty}\pi^{(k)}(s)\big(\mu^{(k)} (s)-r(s-)\big) \\
 \end{eqnarray*}
 \begin{eqnarray*}
\nonumber & & + \sum_{i=0}^{N} \sum_{l=1}^{\infty}\pi^{(l)}_{i}(s)\big(\mu^{(l)}_{i}(s)-r(s-)\big)
+ \frac{1}{2}(\alpha-1)\pi_{0}^{2}(s) \sigma_{0}^{2}(s-)\bigg)\\
\nonumber & & {} + \sum_{j=1}^{N}  \bigg(\big( 1+\pi_{j}(s)\sigma_{j}(s-) \big)^{\alpha} - 1- \alpha  \pi_{j}(s)\sigma_{j}(s-) \bigg) \lambda_{j}(s) \\
\nonumber & & {} +  \int_{\mathbb{R}}  \Bigg(\bigg( 1+\pi_{0}(s)\gamma (s-,x)+ \sum_{k=2}^{\infty}\pi^{(k)}(s)\sigma^{(k)}(s-) \gamma^{k} (s-,x) \bigg)^{\alpha} \\
\nonumber & & {} - 1 - \alpha \Big(\pi_{0}(s)\gamma (s-,x)+ \sum_{k=2}^{\infty}\pi^{(k)}(s)\sigma^{(k)}(s-) \gamma^{k} (s-,x) \Big)\Bigg) \nu (\mathrm{d}x)\\
\nonumber & & {} +\sum_{i=1}^{N} \int_{\mathbb{R}}  \Bigg( \bigg(1+ x \pi_{0}(s)+ \sum_{l=1}^{\infty} x^{l} \pi^{(l)}_{i}(s)\sigma^{(l)}_{i}(s-) \bigg)^{\alpha} - 1 \\
\nonumber & & {}  - \alpha  \Big(x \pi_{0}(s)+ \sum_{l=1}^{\infty} x^{l} \pi^{(l)}_{i}(s)\sigma^{(l)}_{i}(s-)\Big)  \Bigg) \lambda_{i}(s) \eta (\mathrm{d}x)\Bigg] \mathrm{d}s.
\end{eqnarray*}
By direct differentiation with respect to each strategy, this supremum is attained if the strategies satisfy system (\ref{5091}).
Observe that from (\ref{42}) the integrals
\begin{eqnarray*}
\nonumber & & {}  \int_{\mathbb{R}}  \Bigg(\bigg( 1+\pi_{0}(s)\gamma (s-,x)+ \sum_{k=2}^{\infty}\pi^{(k)}(s)\sigma^{(k)}(s-) \gamma^{k} (s-,x) \bigg)^{\alpha} \\
\nonumber & & {} - 1 - \alpha \Big(\pi_{0}(s)\gamma (s-,x)+ \sum_{k=2}^{\infty}\pi^{(k)}(s)\sigma^{(k)}(s-) \gamma^{k} (s-,x) \Big)\Bigg) \nu (\mathrm{d}x)
\end{eqnarray*}
 and (for $i=1,\ldots,N$)
$$ \int_{\mathbb{R}}  \Bigg( \bigg(1+ x \pi_{0}(s)+ \sum_{l=1}^{\infty} x^{l} \pi^{(l)}_{i}(s)\sigma^{(l)}_{i}(s-)\bigg)^{\alpha} - 1  - \alpha  \Big(x \pi_{0}(s)+ \sum_{l=1}^{\infty} x^{l} \pi^{(l)}_{i}(s)\sigma^{(l)}_{i}(s-)\Big)  \Bigg) \lambda_{i}(s) \eta (\mathrm{d}x)$$
 are well-defined. Hence by the Leibniz integral rule we can interchange the derivatives and integrals.

 \halmos

 \begin{remark}
We have not been able to prove that a solution of (\ref{5091}) exists and is unique. However, if we consider a finite market then this solution exists and is unique and must be the optimal portfolio by Kramkov and Schachermayer \cite[Thm. 2.2]{Sch} .
\end{remark}

 \begin{remark}\label{remphen}
If the prices of assets in the Black-Scholes-Merton market are described by processes without jumps (that is, $ \bar{\Pi} (\mathrm{d}t,\mathrm{d}x)= 0 $ and $ \bar{\Pi}_{U}^{i} (\mathrm{d}t,\mathrm{d}x) = 0 $) then we obtain closed-form solutions to the optimal portfolio selection problem (\ref{1136}) for the logarithmic and power utilities (see Zhang et al. \cite{ZSMe}). In addition, the value function in the primary market is the same as in the enlarged market, while in our market this does not occur.
\end{remark}

\section{Optimal portfolio selection in the original market}

\text { } \text { } In this section, we will find conditions for optimal portfolio strategies in the original market, i.e. in the market with one risk-free asset and one share.\\
Let $\tilde{\pi}_{0}$ be the proportion of wealth invested in share $S_{0}$ in the original market. Then the corresponding wealth process, denoted as $R^{\tilde{\pi}_{0}}$, is given by the stochastic differential equation
\begin{eqnarray*}
\dfrac{\mathrm{d}R_{{\tilde{\pi}_{0}}}(t)}{R_{{\tilde{\pi}_{0}}}(t-)} & := & \bigg( r(t-) +\tilde{\pi}_{0}(t)\big(\mu_{0}(t-)-r(t-) \big) \bigg)\mathrm{d}t+\tilde{\pi}_{0}(t) \sigma_{0}(t-)\mathrm{d}W(t)\\
\nonumber & & {} + \tilde{\pi}_{0} (t) \int_{\mathbb{R}} \gamma (t-,x) \bar{\Pi}(\mathrm{d}t,\mathrm{d}x) +\tilde{\pi}_{0} (t) \sum_{i=1}^{N} \int_{\mathbb{R}} x \bar{\Pi}_{U}^{i}(\mathrm{d}t,\mathrm{d}x).
\end{eqnarray*}

Let $\mathcal{A}_{0}$ be the class of admissible portfolio strategies $\tilde{\pi}_{0} $ such that $\tilde{\pi}_{0}$ is predictable, $\{\mathcal{F}_{t}\}$-adaptable and satisfies the condition $\int_{t}^{T} | \tilde{\pi}_{0}(s) |^{2} \mathrm{d}s<\infty, \mathbb{P}- \rm{a. s.}$
Similarly to the definition of the value function in the enlarged market, we define
the value function in the original incomplete market as
\begin{eqnarray*}\label{valfun}
V_{0}(t,z,\textbf{e}_{i}):=\sup\limits_{\tilde{\pi}_{0}\in\mathcal{A}_{0} }\mathbb{E}_{t,z,i}\big[U(R^{\tilde{\pi}_{0}}(T))\big].
\end{eqnarray*}
We assume $\mathbb{E}_{t,z,i}\big[U(R_{\tilde{\pi}_{0}}(T))\big]<\infty$ for $i=1, \ldots,N$.\\

\text { } \text { }  First, we will consider the logarithmic utility function.
We have
\begin{eqnarray*}
\mathbb{E}_{t,z,i}\big[\log R_{\tilde{\pi}_{0}}(T)\big] &= & \log z+ \mathbb{E}_{t,z,i}\int_{t}^{T}\bigg[ r(s-) +\tilde{\pi}_{0}(s)\big(\mu_{0}(s-)-r(s-)\big)-\frac{1}{2}\tilde{\pi}_{0}^{2}(s) \sigma_{0}^{2}(s-)\\
\nonumber & & {} + \int_{\mathbb{R}} \bigg(\log\big(1+\tilde{\pi}_{0}(s)\gamma (s-,x) \big) -\tilde{\pi}_{0}(s)\gamma (s-,x)\bigg)\nu (\mathrm{d}x) \\
\nonumber & & {} + \sum_{i=1}^{N} \int_{\mathbb{R}}\bigg( \log\big(1+x\tilde{\pi}_{0}(s)\big)-x\tilde{\pi}_{0}(s)\bigg)\lambda_{i}(s) \eta (\mathrm{d}x) \bigg]\mathrm{d}s,
\end{eqnarray*}
that is, for the logarithmic utility the value function $V_{0}$ can be written as follows:
\begin{eqnarray*}\label{vfun}
\nonumber V_{0}(t,z,\textbf{e}_{i})&=&  \log z+ \sup\limits_{\tilde{\pi}_{0}\in\mathcal{A}_{0} }\mathbb{E}_{t,z,i}\int_{t}^{T}\bigg[ r(s-) +\tilde{\pi}_{0}(s)\big(\mu_{0}(s-)-r(s-)\big)-\frac{1}{2}\tilde{\pi}_{0}^{2}(s) \sigma_{0}^{2}(s-)\\
 & & {} + \int_{\mathbb{R}} \bigg(\log\big(1+\tilde{\pi}_{0}(s)\gamma (s-,x) \big) -\tilde{\pi}_{0}(s)\gamma (s-,x)\bigg)\nu (\mathrm{d}x) \\
\nonumber & & {} + \sum_{i=1}^{N} \int_{\mathbb{R}}\bigg( \log\big(1+x\tilde{\pi}_{0}(s)\big)-x\tilde{\pi}_{0}(s)\bigg)\lambda_{i}(s) \eta (\mathrm{d}x) \bigg]\mathrm{d}s.
\end{eqnarray*}
From (\ref{42}) the integrals
\begin{eqnarray*}
\int_{\mathbb{R}} \bigg(\log\big(1+\tilde{\pi}_{0}(s)\gamma (s-,x) \big) -\tilde{\pi}_{0}(s)\gamma (s-,x)\bigg)\nu (\mathrm{d}x) \end{eqnarray*}
 and (for $i=1,\ldots,N$)
$$ \int_{\mathbb{R}}\bigg( \log\big(1+x\tilde{\pi}_{0}(s)\big)-x\tilde{\pi}_{0}(s)\bigg)\lambda_{i}(s) \eta (\mathrm{d}x)$$
 are well-defined. Hence by the Leibniz integral rule we can differentiate the above integrals with respect to $\tilde{\pi}_{0}$.
 The above supremum is attained if  $\tilde{\pi}_{0}$ solves the equation
\begin{eqnarray}\label{optima}
& &\mu_{0}(s-)-r(s)- \sigma_{0}^{2}(s-)\tilde{\pi}_{0}(s)+ \int_{\mathbb{R}} \bigg(\frac{\gamma (s-,x)}{1+ \tilde{\pi}_{0}(s)\gamma (s-,x)} -\gamma (s-,x)\bigg)\nu (\mathrm{d}x) \\
\nonumber  & & {} + \sum_{i=1}^{N} \int_{\mathbb{R}}\bigg(\frac{x}{1+x \tilde{ \pi}_{0}(s)} -x \bigg)\lambda_{i}(s) \eta (\mathrm{d}x) =0.
\end{eqnarray}

\text { } \text { }  In the case of the power utility the value function $V_{0}$ can be written as
\begin{eqnarray*}\label{12}
\nonumber  V_0(t,z,\textbf{e}_{i})&=& z^{\alpha} + \sup\limits_{\tilde\pi_0\in\mathcal{A}_0 } \mathbb{E}_{t,z,i} \int_{t}^{T} \Bigg[ \alpha z^{\alpha}\bigg( r(s-)  +\tilde\pi_{0}(s)\big(\mu_{0}(s-)-r(s-)\big)+ \frac{1}{2}(\alpha-1)\tilde\pi_{0}^{2}(s) \sigma_{0}^{2}(s-)\bigg)\\
& & {} +  \int_{\mathbb{R}} z^{\alpha} \Bigg(\bigg( 1+ \tilde\pi_{0}(s)\gamma (s-,x)\bigg)^{\alpha}  - 1 - \alpha \tilde\pi_{0}(s)\gamma (s-,x) \Bigg) \nu (\mathrm{d}x)\\
\nonumber & & {} +\sum_{i=1}^{N} \int_{\mathbb{R}} z^{\alpha}  \Bigg( \bigg(1+x \tilde\pi_{0}(s) \bigg)^{\alpha} - 1 - \alpha x \tilde\pi_{0}(s)  \Bigg) \lambda_{i}(s) \eta (\mathrm{d}x)\Bigg] \mathrm{d}s.
\end{eqnarray*}
Note that from (\ref{42}) the integrals
\begin{eqnarray*}
\int_{\mathbb{R}} z^{\alpha} \Bigg(\bigg( 1+ \tilde\pi_{0}(s)\gamma (s-,x)\bigg)^{\alpha}  - 1 - \alpha \tilde\pi_{0}(s)\gamma (s-,x) \Bigg) \nu (\mathrm{d}x)\end{eqnarray*}
 and (for $i=1,\ldots,N$)
$$ \int_{\mathbb{R}} z^{\alpha}  \Bigg( \bigg(1+x \tilde\pi_{0}(s) \bigg)^{\alpha} - 1 - \alpha x \tilde\pi_{0}(s)  \Bigg) \lambda_{i}(s) \eta (\mathrm{d}x) $$
are well-defined.
By direct differentiation in (\ref{12}) with respect to $\tilde{\pi}_{0}$ we get
\begin{eqnarray}\label{optima2}
& &\mu_{0}(s-)-r(s)-(\alpha-1)\sigma_{0}^{2}(s-)\tilde{\pi}_{0}(s)+ \int_{\mathbb{R}} \bigg(\gamma (s-,x)\big(1+ \tilde{\pi}_{0}(s)\gamma (s-,x)\big)^{\alpha-1} -\gamma (s-,x)\bigg)\nu (\mathrm{d}x) \\
\nonumber  & & {} + \sum_{i=1}^{N} \int_{\mathbb{R}}\bigg(x \big(1+ x \tilde{\pi}_{0}(s)\big)^{\alpha-1} -x \bigg)\lambda_{i}(s) \eta (\mathrm{d}x) =0.
\end{eqnarray}

\begin{lemma}
The solutions of equations (\ref{optima}) and (\ref{optima2}) are optimal strategies for the portfolio selection problem (\ref{valfun}) for the logarithmic and power utilities, respectively.
\end{lemma}

\proof
We will prove that solutions of equations (\ref{optima}) and (\ref{optima2}) are optimal portfolio strategies; their existence was proved by Kramkov and Schachermayer \cite{Sch}.

Let $\tilde{\pi}_{0}^\varepsilon := \tilde{\pi}_{0}+ \varepsilon $  be a perturbed portfolio strategy for $ \varepsilon >0$.

We define the value function $V_{0}^\varepsilon$ related to the strategy  $\tilde{\pi}_{0}^\varepsilon$ (see J.-P. Fouque et al.  \cite{FPS, FSZ}, Mokkhavesa and Atkinson \cite{MA}) as follows:
\begin{eqnarray*}
V_{0}^\varepsilon (t,z,\textbf{e}_{i}):=\sup\limits_{\tilde{\pi}_{0}^\varepsilon \in\mathcal{A}_{0} }\mathbb{E}_{t,z,i}\big[U(R_{\tilde{\pi}_{0}^\varepsilon}(T))\big].
\end{eqnarray*}
In the case of the logarithmic utility,
\begin{eqnarray*}\label{vfun2}
\nonumber V_{0}^\varepsilon (t,z,\textbf{e}_{i})&=&  \log z+ \sup\limits_{\tilde{\pi}_{0}\in\mathcal{A}_{0} }\mathbb{E}_{t,z,i}\int_{t}^{T}\bigg[ r(s-) +(\tilde{\pi}_{0}(s)+\varepsilon)\big(\mu_{0}(s-)-r(s-)\big)-\frac{1}{2}(\tilde{\pi}_{0}(s)+\varepsilon)^{2} \sigma_{0}^{2}(s-)\\
 & & {} + \int_{\mathbb{R}} \bigg(\log\big(1+(\tilde{\pi}_{0}(s)+\varepsilon)\gamma (s-,x) \big) -(\tilde{\pi}_{0}(s)+\varepsilon)\gamma (s-,x)\bigg)\nu (\mathrm{d}x) \\
\nonumber & & {} + \sum_{i=1}^{N} \int_{\mathbb{R}}\bigg( \log\big(1+x(\tilde{\pi}_{0}(s)+\varepsilon)\big)-x(\tilde{\pi}_{0}(s)+\varepsilon)\bigg)\lambda_{i}(s) \eta (\mathrm{d}x) \bigg]\mathrm{d}s,
\end{eqnarray*}
and for the power utility,
\begin{eqnarray*}\label{1}
\nonumber  V_0^\varepsilon (t,z,\textbf{e}_{i})&=& z^{\alpha} + \sup\limits_{\tilde\pi_0\in\mathcal{A}_0 } \mathbb{E}_{t,z,i} \int_{t}^{T} \Bigg[ \alpha z^{\alpha}\bigg( r(s-)  +(\tilde\pi_{0}(s)+\varepsilon)\big(\mu_{0}(s-)-r(s-)\big)\\
\end{eqnarray*}
\begin{eqnarray*}
& & {} + \frac{1}{2}(\alpha-1)(\tilde\pi_{0}(s)+\varepsilon)^{2} \sigma_{0}^{2}(s-)\bigg)\\
\nonumber  & & {}+  \int_{\mathbb{R}} z^{\alpha} \Bigg(\bigg( 1+ (\tilde\pi_{0}(s)+\varepsilon )\gamma (s-,x)\bigg)^{\alpha}  - 1 - \alpha (\tilde\pi_{0}(s)+\varepsilon )\gamma (s-,x) \Bigg) \nu (\mathrm{d}x)\\
\nonumber & & {} +\sum_{i=1}^{N} \int_{\mathbb{R}} z^{\alpha}  \Bigg( \bigg(1+x (\tilde\pi_{0}(s)+\varepsilon ) \bigg)^{\alpha} - 1 - \alpha x (\tilde\pi_{0}(s)+\varepsilon )  \Bigg) \lambda_{i}(s) \eta (\mathrm{d}x)\Bigg] \mathrm{d}s.
\end{eqnarray*}
Note that $\tilde\pi_0$ is a portfolio strategy that maximizes the value function, so $\frac{\partial}{\partial \varepsilon} V_{0}^\varepsilon  \bigg\arrowvert _{ \varepsilon=0 } =0$. Calculating this derivative for $V_{0}^\varepsilon$ in both cases, we get  equations (\ref{optima}) and (\ref{optima2}).

Thus the optimal portfolio strategies solve these equations. The existence and uniqueness of the optimal strategies follows from Kramkov and Schachermayer \cite{Sch}. In that paper the main assumption concerns the utility function, which has to have asymptotic elasticity strictly less than $1$, that is,
\begin{equation*}\label{518}
\overline{\lim_{z \to \infty}} \ \  \dfrac{zU'(z)}{U(z)}<1.
\end{equation*}
Note that the power and logarithmic utilities satisfy this condition.
 \halmos

\begin{remark}
In a general semimartingale market model, Goll and Kallsen \cite{GK1, GK2}  obtained the optimal solution explicitly in terms of  semimartingale characteristics of the price process for the logarithmic utility.

\end{remark}

\end{document}